%% file: 00_main.tex
\documentclass[sigconf]{acmart}

\usepackage{amsthm}
\usepackage{amsmath}
\usepackage{times}
\usepackage{kotex}
\usepackage{soul}
\usepackage{xcolor}
\usepackage{url}
\usepackage{enumitem}
\usepackage{comment}
\usepackage{threeparttable}
\usepackage[switch]{lineno}
\usepackage{lipsum}
\usepackage[linesnumbered,algoruled,boxed,lined,noend]{algorithm2e}
\usepackage{graphicx,subcaption}
\usepackage{fancyhdr} 
\usepackage[utf8]{inputenc}
\usepackage{multirow}
\usepackage[normalem]{ulem}
\usepackage{multicol}
\usepackage{makecell}
\usepackage{cancel}
\usepackage{cellspace}
\usepackage{fontawesome}
\usepackage{microtype}
\usepackage{float}
\usepackage{placeins} 
\usepackage{stfloats}

\newcommand{\spara}[1]{\smallskip\noindent{\bf #1}}

\newtheorem{definition}{Definition}
\newtheorem{property}{Property}

\newtheorem{example}{Example}

\newcommand{\EPA}{\textsf{EPA}}

\newcommand{\BCA}{\textsf{BCA}}

\AtBeginDocument{%
  }

\setcopyright{none}
\copyrightyear{2018}
\acmYear{2018}
\acmDOI{XXXXXXX.XXXXXXX}
\acmConference[Conference acronym 'XX]{Make sure to enter the correct
  conference title from your rights confirmation email}{June 03--05,
  2018}{Woodstock, NY}

\acmISBN{978-1-4503-XXXX-X/2018/06}

\begin{document}
\fancyfoot{}

\makeatletter
\def\fps@figure{tbp}
\def\fps@figure*{tbp} 
\makeatother

\title{Uncovering High-Order Cohesive Structures: \\Efficient $(k,g)$-Core Computation and Decomposition for Large Hypergraphs}

\author{Dahee Kim$^\S$, Hyewon Kim$^\S$, Song Kim$^\S$, Minseok Kim$^\S$, \\ Junghoon Kim$^{\S}$\footnotemark, Yeon-Chang Lee$^\S$, Sungsu Lim$^\dagger$}
\affiliation{%
  \institution{$^\S$Ulsan National Institute of Science \& Technology, Ulsan, Republic of Korea}
  \city{$^\dagger$Chungnam National University, Daejeon, Republic of Korea}
  \country{}}
  
\email{{dahee, hyewon.kim, song.kim, end423, junghoon.kim, yeonchang}@unist.ac.kr, sungsu@cnu.ac.kr}

\renewcommand{\shortauthors}{Kim et al.}

\begin{abstract}
Hypergraphs, increasingly utilised to model complex and diverse relationships in modern networks, have gained significant attention for representing intricate higher-order interactions. Among various challenges, cohesive subgraph discovery is one of the fundamental problems and offers deep insights into these structures, yet the task of selecting appropriate parameters is an open question. To address this question, we aim to design an efficient indexing structure to retrieve cohesive subgraphs in an online manner. The main idea is to enable the discovery of corresponding structures within a reasonable time without the need for exhaustive graph traversals. Our method enables faster and more effective retrieval of cohesive structures, which supports decision-making in applications that require online analysis of large-scale hypergraphs. Through extensive experiments on real-world networks, we demonstrate the superiority of our proposed indexing technique.
\end{abstract}

\maketitle

\def\thefootnote{*}
\footnotetext{Corresponding author}
\def\thefootnote{\arabic{footnote}}  

\input{01_introduction}
\input{02_problem_statement}

\input{03_kgcore}

\input{04_decomposition}

\input{05_experiments}

\input{06_related_work}

\input{07_conclusion}

\spara{Acknowledgment.}
This work was supported by Institute of Information \& communications Technology Planning \& Evaluation(IITP) grant funded by the Korea government(MSIT) (No. RS-2020-II201336, Artificial Intelligence Graduate School Program(UNIST)) and by the National Research Foundation of Korea (NRF) grant funded by the Korea government (MSIT) (No.RS-2023-00214065, RS-2025-00523578) and Basic Science Research Program through the
National Research Foundation of Korea(NRF) funded by the Ministry of Education(No.RS-2024-00354951).
This work was in part supported by Institute of Information \& communications Technology Planning \& Evaluation(IITP) under the Leading Generative AI Human Resources Development(IITP-2025-RS-2024-00360227) grant funded by the Korea government(MSIT).

\bibliographystyle{ACM-Reference-Format}
\bibliography{98_bib}

\clearpage

\appendix
\input{80_appendix}


\end{document}

%% file: 01_introduction.tex
\section{INTRODUCTION}

Graphs are a fundamental tool for modelling diverse relationships in the real-world, such as social sciences, biology, and transaction~\cite{harary1953graph,mason2007graph,jack2013transaction}. Graph analysis enables revealing the structural properties of network by identifying interconnected entities and meaningful patterns~\cite{wellman1983network}.
One of the key tasks in graph analysis is to identify cohesive subgraphs, which are subsets of nodes that exhibit stronger internal connectivity than their connections to the rest of the graph~\cite{benson2018simplicial, malliaros2020core}. These cohesive subgraphs are useful in numerous applications, including community detection in social networks~\cite{cohen2008trusses}, identification of functional modules in biological systems~\cite{salem2012discovering}, and analysis of user groups in e-commerce~\cite{xia2021self}.

Numerous approaches have been widely used in traditional graphs to identify cohesive subgraphs, such as the $k$-core~\cite{malliaros2020core}, modularity-based clustering~\cite{girvan2002community}, and spectral partitioning~\cite{gunnemann2013spectral}. However, existing approaches primarily handle pairwise relationships, relying on indirect representations of higher-order structures such as motifs and cliques, for relationships between multiple entities. 
While these methods provide a partial representation of these complex structures, literatures~\cite{gao2022hgnn+,li2023hypergraph,xia2021self} suggest that explicitly incorporating higher-order relationships can significantly enhance performance in downstream tasks, including co-authoring analysis, co-purchasing predictions, and group participation modelling.

To better capture higher-order structures beyond pairwise interactions,
hypergraphs have emerged as a more flexible structure where a single hyperedge can connect multiple nodes simultaneously~\cite{berge1984hypergraphs}.
%
%
Various models~\cite{nbrkcore, bu2023hypercore, leng2013m} have been proposed for discovering cohesive subhypergraphs, where a subhypergraph refers to the substructure of a hypergraph formed by a subset of nodes and the hyperedges that contain those nodes.
The $k$-hypercore~\cite{leng2013m} represents the first extension of the $k$-core~\cite{seidman1983network} to hypergraphs. However, $k$-hypercore only consider neighbour constraint, it often produces overly broad results, including nodes with weak connectivity with each other.

Subsequently, the $(k,d)$-core~\cite{nbrkcore} was introduced, which incorporates a \textit{degree} constraint at the individual node level to identify more cohesive subhypergraphs. However, this model assumes that if any node in a hyperedge is removed, the hyperedge containing the removed node must also be discarded. This assumption does not always hold in real-world scenarios, where hyperedge can remain meaningful even if some nodes are removed, leading to unnecessary information loss~\cite{bu2023hypercore}. 
Furthermore, the $(k,d)$-core may produce subhypergraphs with weak interconnectivity since it focuses primarily on the strength of individual node, which may miss important patterns of connectivity between neighbour nodes.
For example, if nodes are frequently involved together in multiple groups (\textit{i.e.}, hyperedges), their relationship may still be significant, even when their individual participation (\textit{i.e.}, degree) is low.
This \textit{co-occurrence pattern} can revel latent structural dependencies that are not captured by degree-based constraints alone.

%

Therefore, this paper aims to incorporate the concept of ``co-occurrence,'' which accounts for the frequency of repeated interactions between nodes in hypergraphs. 
These interactions reflect the strength of group relationships, as demonstrated in Example~\ref{ex:motivation}, enabling the identification of stronger and more structurally cohesive subhypergraphs. 


\begin{figure}[t]
    \centering
    \includegraphics[width=0.79\linewidth]{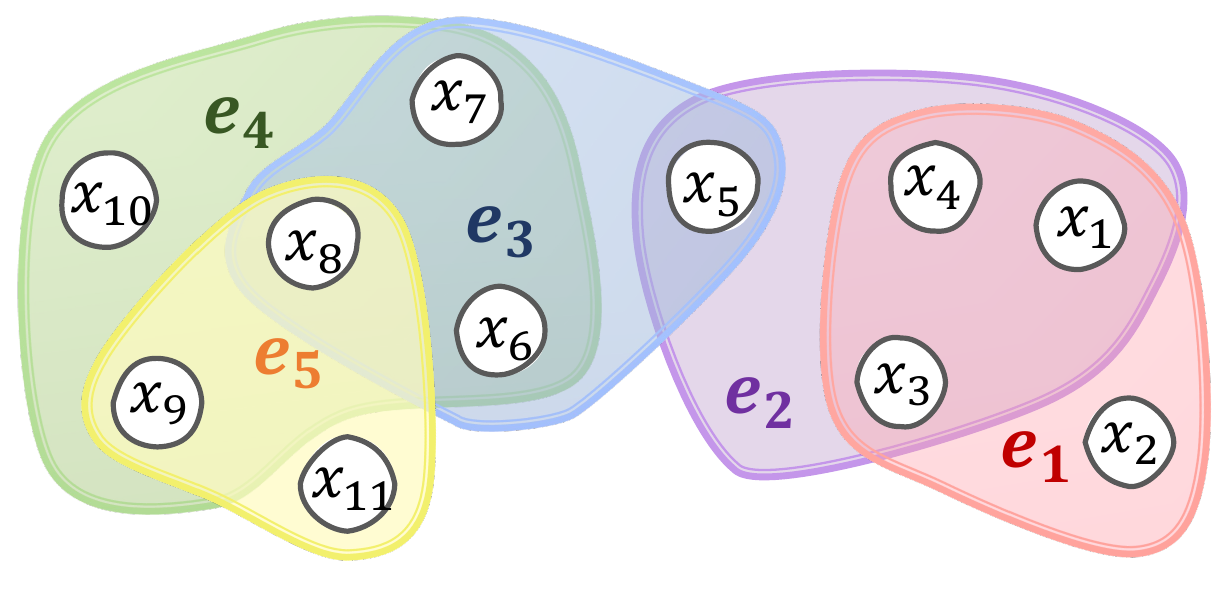}
    \vspace{-0.3cm}
    \caption{Toy example}
    \vspace{-0.4cm}
    \label{fig:example}
\end{figure}

\begin{example}
\label{ex:motivation}
Consider a research-collaboration platform where each hyperedge denotes a team project, and each node represents a researcher as shown in Figure~\ref{fig:example}. Our goal is to discover subgroups of researchers who collaborate frequently, thus enabling recommendation of future collaborations. Suppose every researcher in the subgroup is required to have participated in at least two team projects. If we rely solely on the number of team projects that each researcher participates in (\textit{i.e.}, degree-based approach), we would select a subgroup that includes members $x_1, x_3, x_4,x_5, x_6, x_ 8$, and $x_9$. Although $x_1$ and $x_8$ both meet the participation requirement, they collaborate with completely different sets of researchers. In contrast, if we account for co-occurrence, we can identify two subgroups: $\{x_1, x_3, x_4\}$, and $\{x_6, x_7,x_8\}$. Members of each subgroup consistently collaborate with each other across all of their projects. Such co-occurrence patterns capture groups that are strongly interconnected, whereas relying on degree may lead to identifying groups with weaker interconnections.
\end{example}

Motivated by this observation, we propose the $(k,g)$-core, a novel model for discovering cohesive subhypergraphs by incorporating co-occurrence. Compared with the $(k,d)$-core~\cite{nbrkcore} that focuses on individual node strength, the $(k,g)$-core captures more strongly connected groups by considering co-occurrence, a condition that requires node pairs to frequently appear together in multiple hyperedges.
%
%
Under the $(k,g)$-core model, we design two key algorithms: (i) the Efficient Peeling Algorithm (Section~\ref{sec:kgcore_comp}), which ensures memory efficiency and scalability by maintaining only essential information during $(k,g)$-core computation, and (ii) the Bucket-based Coreness Algorithm (Section~\ref{sec:kgcore_decomp}), which decomposes a hypergraph into all possible $(k,g)$-cores.

\spara{Application.} The $(k,g)$-core model can be applied in a variety of domains, some notable applications are as follows: 

\begin{itemize}[leftmargin=*] 
    \item \textbf{\textit{Team formation}}: Team formation is an important task in organizational and research collaboration settings~\cite{li2010team}. Typically, each hyperedge represents a group of researchers who collaborate on publications or projects. The primary advantage of using $(k,g)$-cores is that it prevents teams from being formed by researchers with high individual engagement but low shared collaboration history. This helps to avoid the formation of teams with weak internal cohesion, resulting in subgroups of members who demonstrate stable and reliable collaborative relationships.
    \item \textbf{\textit{Market basket analysis}}: 
    Market basket analysis is a widely used technique for marketers to create customer-oriented strategies~\cite{cciccekli2021market}. Basket data can be modelled as a hypergraph by representing each basket as a hyperedge and items as a node. Traditional market basket analysis\cite{unvan2021market,saputra2023market} focuses on the most frequently sold items, often relying on global frequency thresholds. By applying the $(k,g)$-core, however, it can identify a set of items that are repeatedly purchased by multiple customers, thus revealing valuable cross-selling patterns even if the total sales of individual items are not particularly high.

\end{itemize}


\spara{Contributions.} 
The key contributions of this paper are summarised as follows:
\begin{itemize}[leftmargin=*] 
\item \textbf{A new cohesive subhypergraph model.}
We propose the $(k,g)$-core, a novel model that efficiently identifies cohesive subhypergraphs while effectively capturing co-occurrence patterns in hypergraphs.

\item \textbf{Efficient algorithms.} 
We introduce two algorithms: the Efficient Peeling Algorithm (\EPA) for $(k,g)$-core computation, which reduces unnecessary space usage, and the Bucket-based Coreness Algorithm (\BCA) for $(k,g)$-core decomposition.

\item \textbf{Extensive experimental evaluation.} We evaluate our algorithms on eight real-world hypergraph datasets, demonstrating their scalability and effectiveness in identifying cohesive subgraphs. The $(k,g)$-core decomposition further provides structural insights into cohesive subgraphs across real-world datasets.
\end{itemize}

\spara{Extension of Previous Work.}
An earlier version of this work was introduced as a short paper in ACM CIKM 2023~\cite{kgcore}, where we introduced the $(k,g)$-core concept, a straightforward algorithm for its computation, and preliminary experiments.
In this extended version, we present several key advancements:
\begin{enumerate}[leftmargin=*]
\item We enrich the problem motivation with improved examples, a more detailed background discussion, and strengthened theoretical foundations;
\item We introduce the \textit{Efficient Peeling Algorithm} (\EPA), which significantly reduces space complexity while maintaining the same time complexity as the greedy algorithm in~\cite{kgcore};
\item We extend our focus beyond $(k,g)$-core computation to address the decomposition of $(k,g)$-core, for which we propose the \textit{Bucket-based Coreness Algorithm} (\BCA); and
\item We significantly expand our experimental evaluation, incorporating large-scale real-world datasets and additional experiments to validate the performance and scalability of our approach.
\end{enumerate}


\spara{Paper Structure.} 
The remainder of this paper is structured as follows: Section~\ref{sec:problemstatement} introduces key concepts and notations for hypergraph-based subgraph models, leading to the formal definition of the $(k,g)$-core. Section~\ref{sec:kgcore_comp} presents the Efficient Peeling Algorithm (\EPA), a top-down heuristic approach for the computation of $(k,g)$-core. Section~\ref{sec:kgcore_decomp} proposes the Bucket-based Coreness Algorithm (\BCA) of $(k,g)$-core decomposition, providing theoretical insights and algorithmic implementation. Section~\ref{sec:experiments} evaluates the two key algorithms
on real-world hypergraphs to demonstrate their superiority. Section~\ref{sec:relatedwork} reviews cohesive subgraph models in hypergraphs and traditional graphs. Finally, Section~\ref{sec:conclusion} summarises findings and future research directions.

%% file: 02_problem_statement.tex
\section{PROBLEM STATEMENT}
\label{sec:problemstatement}


\begin{table}[t]
\centering
\small
\caption{Notations}
\label{tab:notations}
\begin{tabular}{c||l}
\hline \hline
\textbf{Notation} & \textbf{Description}          \\ \hline \hline
$G=(V, E)$       & Hypergraph   \\ \hline
$G[V']$ & Induced subhypergraph by $V'\subseteq V$ \\ \hline
$\mathcal{E}(v)$ & Set of hyperedges containing node $v$ in $G$ \\ \hline
$node(e)$& Set of nodes in hyperedge $e$\\ \hline
\multirow{2}{*}{$deg(v)$} & Degree of node $v$ \\ & \textit{i.e.}, the count of hyperedges containing $v$ \\ \hline
\multirow{2}{*}{$|e|$} & Cardinality of hyperedge $e$ \\ & \textit{i.e.}, the number of nodes in $e$ \\ \hline
\multirow{2}{*}{$s(u,v)$} & Support of node pair $u,v$ \\ & \textit{i.e.}, the number of hyperedges involving $u, v$  \\ \hline
\multirow{2}{*}{$N_g(v)$} & Set of $g$-neighbours of node $v$ \\  & \textit{i.e.}, nodes that share at least $g$ hyperedges with $v$ \\ \hline \hline
\end{tabular}
\end{table}

This section introduces the fundamental notations and concepts essential for this study. 
The main notations used throughout the paper are summarised in Table~\ref{tab:notations}.

To clarify the terminology in the hypergraph context, we define several key concepts. 
A hypergraph is denoted as $G = (V, E)$, where $V$ is the set of nodes $E$ is the set of hyperedges. In this work, we focus on undirected and unweighted hypergraphs. 
An \emph{induced subhypergraph} $G[V']$ contains all nodes in $V' \subseteq V$ and only the hyperedges in $E$ that connect at least two nodes in $V'$. The \emph{cardinality} $|e|$ of a hyperedge $e \in E$ refers to the number of nodes it contains. The \emph{degree} of a node $v \in V$, denoted as $deg(v)$, is the number of hyperedges that include $v$. 

To capture the strength of connections between nodes in a hypergraph, we introduce the concept of support.


\begin{definition}[\underline{\textbf{Support}}]
    Given a hypergraph $G=(V,E)$, the support value $s(u,v)$ with two nodes $u,v\in V$ in $G$ is defined as the number of hyperedges in which both nodes co-occur, \textit{i.e.},  $|\{e\in E: u\in e \wedge v\in e\}|$. 
\end{definition}


\begin{definition}[\underline{\textbf{$g$-neighbour}}]
Given a hypergraph $G = (V, E)$, a node $u \in V$, and a support threshold $g$, a node $v \in V$ is called a $g$-neighbour of $u$ in $G$ if the support value between $u$ and $v$ is greater than or equal to $g$, \textit{i.e.}, $s(u, v) \geq g$.
\end{definition}

We denote the set of $g$-neighbours of a node $u$ as $N_g(u)$. This concept captures both the presence of connections between nodes and their strength, as measured by the number of shared hyperedges. Building on the notion of $g$-neighbours, we introduce the concept of the $(k,g)$-core in hypergraphs.

\begin{definition}[\underline{$\boldsymbol{(k,g)}$\textbf{-core}}~\cite{kgcore}] Given a hypergraph $G = (V, E)$ a neighbour size threshold $k$, and a support threshold $g$, the $(k,g)$-core, denoted as $C_{k, g}$, is the maximal set of nodes where each node has at least $k$ neighbour nodes as its $g$-neighbours within $G[C_{k, g}]$.
\end{definition}

We introduce two key properties of the $(k, g)$-core that characterise its structure: uniqueness and containment.

\begin{property}
\underline{Uniqueness of $(k,g)$-core}: The $(k,g)$-core is unique in that it represents the maximal subset of nodes satisfying the given $k$ and $g$ constraint within the hypergraph.
\end{property}

\begin{property}
\underline{Containment of $(k,g)$-core}: The $(k,g)$-core has a  hierarchical structure, meaning that both the $(k+1,g)$-core and the $(k,g+1)$-core are contained within the $(k,g)$-core.
\end{property}

The proof is omitted as it follows directly from the definition. Intuitively, if two nodes appear together in five hyperedges, they necessarily appear together in at least four. Similarly, if a node has at least four $g$-neighbours, it must also have at least three. 

\begin{example}
Consider a hypergraph with $11$ nodes and $5$ hyperedges, as shown in Figure~\ref{fig:example}. For $k=2$ and $g=2$, nodes $\{x_1, x_3, x_4\}$ form a cohesive group with a support value of $2$, as they co-occur in hyperedges $e_1$ and $e_2$. Similarly, nodes $\{x_6, x_7, x_8\}$ share a support value of $2$ by appearing together in $e_3$ and $e_4$. Each of these nodes has at least two $g$-neighbours, satisfying the $(2,2)$-core conditions. In contrast, nodes $\{x_2, x_5, x_{10}, x_{11}\}$ are not part of the $(2,2)$-core as they do not meet the $g$ constraint, having a support value of at most $1$ with their neighbours. Node $x_9$ is also not included since, despite having a support value of $2$ with $x_8$, it has only one $g$-neighbour, violating the $k$ constraint. Hence, the $(2,2)$-core is $\{x_1, x_3, x_4, x_6, x_7, x_8\}$.
\end{example}

%% file: 03_kgcore.tex
\section{$(k,g)$-CORE COMPUTATION}
\label{sec:kgcore_comp}

This section introduces the Efficient Peeling Algorithm (\EPA), an optimised approach to $(k,g)$-core computation designed to address the scalability limitations of an existing approach. In particular, a naive baseline algorithm is described in ~\ref{appendix:naive} for reference. 
%

The naive approach maintains an explicit $g$-neighbour set for every node and continuously updates the pairwise co-occurrence counts. While effective in small or sparse settings, this method can require $O(|V|^2)$ memory on dense hypergraphs, making it impractical to handle large-sized datasets.

In contrast, {\EPA} only tracks the number of $g$-neighbours for each node rather than storing the full co-occurrence information. This reduces the memory requirement to $O(|V|)$ while preserving the correctness of the resulting $(k,g)$-core. Consequently, {\EPA} significantly improves memory efficiency and enhances scalability without compromising accuracy.

A detailed comparison of the memory consumption between the naive algorithm and {\EPA} is provided in Appendix~\ref{appendix:naive_memory}.

\begin{algorithm}[t]
\small
\SetAlgoLined
\SetKwData{break}{break}
\SetKwData{AND}{AND}
\SetKwData{false}{false}
\SetKwData{true}{true}
\SetKwData{del}{delete}
\SetKwFunction{queueInit}{queue}
\SetKwFunction{emptyc}{isEmpty}
\SetKwFunction{pop}{dequeue}
\SetKwFunction{contain}{contain}
\SetKwFunction{push}{enqueue}
\SetKwFunction{getOccurMap}{getOccurMap}
\SetKwFunction{getNbrMap}{getNbrMap}
\SetKwFunction{getKeys}{getKeys}
\KwIn{Hypergraph $G=(V,E)$, parameters $k$ and $g$}
\KwOut{The $(k, g)$-core of $G$}
$C \leftarrow \{\};$ \tcp{Store the $g$-neighbour size for each node}  
$H \leftarrow V;$ \tcp{Candidate nodes for $(k,g)$-core} 
$VQ \leftarrow$ \queueInit{}; \tcp{Store nodes that do not satisfy the constraints} 
\For{$v \in H$}{
    $N_g (v) \leftarrow$ \getNbrMap{$H, E, v, g$}\;
    $C[v] \leftarrow \left|N_g (v)\right|$\;
    \If{$C[v] < k$}{
        $VQ$.\push{$v$}\;
    }
}

\While{$VQ$.\emptyc{} $\neq$ \true}{
    $v \leftarrow VQ$.\pop{}\; 
    $N_g (v) \leftarrow$ \getNbrMap{$H, E, v, g$}\;
    $H \leftarrow H \setminus \{v\}$\; 
    \For{$w \in$ \getKeys{$N_g (v)$}}{
        \If{$VQ$.\contain{$w$} $\neq$ \true}{
            $C[w] \leftarrow C[w]-1$\;
            \If{$C[w]<k$}{
                $VQ$.\push{$w$}\;
            }
        }
    }
    \del $C[v]$\;
}
\Return{$H$}
\caption{\mbox{Efficient peeling algorithm (\EPA)}}
\label{alg:EPA}
\end{algorithm}

\begin{figure*}[t]
    \centering
    \includegraphics[width=0.99\linewidth]{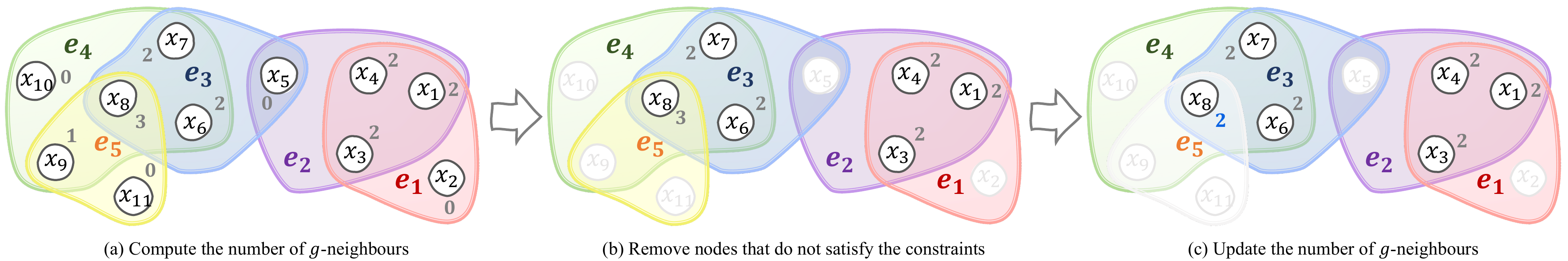}
    \caption{Computation process of $g$-neighbour counts in Figure~\ref{fig:example}}
    \label{fig:ex3}
\end{figure*}

\spara{Pseudo Description.} 
Algorithm~\ref{alg:EPA} presents the overall procedure of {\EPA}. Initially, for each node, the size of its $g$-neighbour is computed and recorded (Lines 4-6). Nodes that do not satisfy the $(k, g)$-core criteria, \textit{i.e.}, those with fewer than $k$ $g$-neighbours, are identified and marked for removal (Lines 7-8).  
The algorithm proceeds iteratively, deleting the nodes marked for removal at each step (Lines 9-12). This removal impacts the $g$-neighbours of the deleted node, and the number of $g$-neighbours for each affected node is updated (Lines 13-15). If the updated $g$-neighbour count for a node falls below $k$, that node is also marked for removal in subsequent iterations (Lines 16-17).  
This iterative peeling process continues until all remaining nodes satisfy the $(k, g)$-core conditions (Line 9). The final set of nodes constitutes the $(k,g)$-core, where each node has at least $k$ $g$-neighbours.

\begin{example}
Consider the hypergraph shown in Figure~\ref{fig:example} with parameters $k=2$ and $g=2$. The details of the {\EPA} process are illustrated in Figure~\ref{fig:ex3}, where each node is annotated with its corresponding $g$-neighbour count in grey.
Initially, the $g$-neighbours of each node are calculated, and only their count is stored, as illustrated in Figure~\ref{fig:ex3}(a).
As shown in Figure~\ref{fig:ex3}(b), after calculating the $g$-neighbours, nodes that do not satisfy the $k$ constraint, such as $x_2$, $x_5$, $x_9$, $x_{10}$, and $x_{11}$, are removed. As each node is removed, the $g$-neighbour count of its $g$-neighbours is updated. Since the only $g$-neighbour of $x_9$ is $\{x_8\}$, removing $x_9$ reduces the $g$-neighbour count of $x_8$. After the update, $x_8$ still has a $g$-neighbour count of $2$, satisfying the constraints and remaining in the $(k,g)$-core. Finally, {\EPA} returns $\{x_1, x_3, x_4, x_6, x_7, x_8\}$. This update process and the final result of the $(k,g)$-core are illustrated in Figure~\ref{fig:ex3}(c).
\end{example}

\spara{Time Complexity.}
The time complexity of {\EPA} is $O(|e^*| \cdot \mathcal{D})$, where $\mathcal{D}$ denotes the total sum of the degree of all nodes in the hypergraph, and $|e^*|$ denotes the maximum cardinality among all hyperedges. To compute the $g$-neighbours of a single node $v$, the algorithm examines each hyperedge containing $v$—there are $deg(v)$ such hyperedges. For each of these, it iterates through up to $|e^*|$ nodes to identify those that co-occur with $v$, resulting in a time complexity of  $O(|e^*| \cdot deg(v))$. Considering all nodes in the hypergraph, the total time complexity becomes $O(|e^*| \cdot \sum_{v \in V} deg(v)) = O(|e^*| \cdot \mathcal{D})$.

\spara{Space Complexity.}
The space complexity of {\EPA} is $O(|V|)$, as it stores only the size of $g$-neighbours. This space complexity makes the {\EPA} suitable for large-sized hypergraphs.

%% file: 04_decomposition.tex
\section{$(k,g)$-CORE DECOMPOSITION}
\label{sec:kgcore_decomp}

In this section, we introduce the concept of coreness for the $(k,g)$-core and present a decomposition method that computes $(k,g)$-cores for all combinations of $k$ and $g$ values. 
Due to the dual constraints imposed by the $(k,g)$-core definition—namely the neighbour threshold ($k$) and the co-occurrence support ($g$)—existing core decomposition techniques~\cite{batagelj2003m, leng2013m} cannot be directly applied. To address this challenge, we propose the Bucket-based Coreness Algorithm (\BCA), a memory-efficient and scalable algorithm that systematically explores and extracts the full hierarchy of $(k,g)$-cores.
In Section~\ref{sec:kgcore_comp}, we address the $(k,g)$-core for a given $(k,g)$ pair; in this section, we reveal the entire hierarchical structure of the network by decomposing it into all possible $(k,g)$-cores. This hierarchical structure enables a deeper analysis of the network to identify various levels of cohesion.

\subsection{\texorpdfstring{Coreness of the $(k,g)$-core}{Coreness of the $(k,g)$-core}}

As mentioned in Section~\ref{sec:problemstatement}, the $(k,g)$-core has uniqueness and containment properties, which allow us to define a concept of node coreness. The coreness of a node reflects the maximum level of $(k,g)$-core that the node can belong to, thereby indicating the intensity of engagement of each node within the network~\cite{linghu2020global,zhang2023quantifying}.
We first introduce the concepts of $k$-coreness and $g$-coreness, which capture, respectively, the highest $g$ and $k$ values for which a node remains in the corresponding $(k,g)$-core. Based on these measures, we define the $(k,g)$-coreness as a comprehensive measure.


\begin{definition}
    (\underline{$\boldsymbol{k}$\textbf{-coreness}}). Given a hypergraph $G=(V,E)$ and a positive integer $k$, the $k$-coreness of a node $v\in V$
    is defined as the maximum integer $g'$ for which $v$ belongs to the $(k,g')$-core but not to the $(k,g'+1)$-core.
\end{definition}
\begin{definition}
    (\underline{$\boldsymbol{g}$\textbf{-coreness}}). Given a hypergraph $G=(V,E)$ and a positive integer $g$, the $g$-coreness of a node $v\in V$
    is defined as the maximum integer $k'$ for which $v$ belongs to the $(k',g)$-core but not to the $(k'+1,g)$-core.
\end{definition}

The $k$-coreness indicates maximal co-occurrence frequency that a node can have and the $g$-coreness represents the highest levels of neighbour connectivity. These measures help quantify the extent to which a node is included in the $(k,g)$-core structure. They serve as key metrics in efficiently constructing the $(k,g)$-core decomposition, providing insight into the cohesive properties of nodes within the network. Through the integration of $k$-coreness and $g$-coreness, we introduce the $(k,g)$-coreness, which provides a comprehensive measure of node participation.

\begin{definition}
(\underline{$\boldsymbol{(k,g)}$\textbf{-coreness}}). Given a hypergraph $G=(V,E)$ and a node $v\in V$, the $(k, g)$-coreness of a node $v$ is the maximal $(k, g)$ pairs such that $v$ is in the $(k,g)$-core but not in the $(k', g)$-core or the $(k, g')$-core, where $k' > k$ and $g' > g$.
\end{definition}

\begin{table}[t]
\centering
\caption{($k,g$)-core and Coreness in Figure~\ref{fig:example}}
\vspace{-0.3cm}
\label{tab:core_coreness}
\begin{subtable}{0.25\textwidth}
\footnotesize
\centering
\subcaption{($k,g$)-core}
\label{tab:cores}
\begin{tabular}{c||c}
\hline \hline
$\boldsymbol{(k,g)}$ & $\boldsymbol{(k,g)}$\textbf{-core}  \\
\hline \hline
$(1,1)$ & $\{x_1,\cdots,x_{11}\}$  \\
\hline
$(2,1)$ & $\{x_1,\cdots,x_{11}\}$  \\
\hline
$(3,1)$ & $\{x_1,\cdots,x_{10}\}$  \\
\hline
$(4,1)$ & $\{x_6,\cdots,x_{10}\}$  \\
\hline 
$(1,2)$ & $\{x_1,x_3,x_4,x_6,\cdots,x_9\}$  \\
\hline
$(2,2)$ & $\{x_1,x_3,x_4,x_6,x_7,x_8\}$ \\ \hline \hline
\end{tabular}
\end{subtable}%
\begin{subtable}{0.25\textwidth}
\footnotesize
\subcaption{$(k,g)$-coreness}
\label{tab:coreness}
\centering
\begin{tabular}{c||c}
\hline \hline
\textbf{Nodes} & $\boldsymbol{(k,g)}$\textbf{-coreness} \\
\hline \hline
$v_1, v_3, v_4$ & $(3,1),(2,2)$ \\ \hline 
$v_2, v_5$ & $(3,1)$ \\ \hline
$v_6, v_7, v_8$ & $(4,1),(2,2)$ \\ \hline
$v_9$ & $(4,1),(1,2)$ \\ \hline
$v_{10}$ & $(4,1)$ \\ \hline
$v_{11}$ & $(2,1)$ \\ \hline \hline
\end{tabular}
\end{subtable}
\end{table}

The $(k,g)$-coreness reflects the significance of nodes within the network and their connectivity to neighbours by considering two distinct parameters. Note that a node can have multiple $(k,g)$-coreness values, as long as each is maximal.

\begin{example}
    Consider the hypergraph shown in Figure~\ref{fig:example}. The corresponding $(k,g)$-cores and $(k,g)$-coreness values of each node are described in Table~\ref{tab:cores} and Table~\ref{tab:coreness}, respectively. Nodes $x_1, x_2,x_3,x_4,$ and $x_5$ belong to the $(1,1)$-core, $(2,1)$-core, and the $(3,1)$-core. However, since $(k,g)$-coreness is defined by the maximal $(k,g)$ pairs, they are assigned only to the $(3,1)$-coreness. Node $v_6$ holds multiple coreness values of $(4,1)$ and $(2,2)$, as it remains in these corresponding $(k,g)$-cores but does not qualify for any stricter $(k',g)$-core or $(k,g')$-core with $k'> k$ or $g'> g$.
\end{example}

This leads to the uniqueness property of $(k, g)$-coreness, which is as follows:

\begin{property}
\underline{Uniqueness of $(k,g)$-coreness}: The $(k,g)$-coreness is unique in that it corresponds to the highest pair of $(k,g)$ values for which the node is included in the $(k,g)$-core but not in any higher $(k',g)$-core or $(k,g')$-core, where $k'>k$ and $g'>g$.
\end{property}

\subsection{\texorpdfstring{Bucket-based decomposition algorithm}{Bucket-based decomposition algorithm}} This section introduces the Bucket-based Coreness Algorithm (\BCA), a decomposition method that computes the $(k,g)$-coreness of each node by identifying the highest $(k,g)$ values for which it remains in a $(k,g)$-core. The process of {\BCA} involves two steps: (1) enumerating all possible $(k,g)$-cores and (2) removing duplicates by leveraging the hierarchical properties of the $(k,g)$-core. 

\begin{algorithm}[h]
\footnotesize
\SetAlgoLined
\SetKwData{false}{false}
\SetKwData{true}{true}
\SetKwData{del}{delete}
\SetKwData{break}{break}
\SetKwFunction{queueInit}{queue}
\SetKwFunction{emptyc}{isEmpty}
\SetKwFunction{pop}{dequeue}
\SetKwFunction{push}{enqueue}
\SetKwFunction{remove}{remove}
\SetKwFunction{pushAll}{enqueueAll}
\SetKwFunction{size}{size}
\SetKwFunction{getNbrMap}{getNbrMap}
\SetKwFunction{Deduplication}{Deduplication}
\SetKwFunction{buildBuckets}{buildBuckets}
\SetKwFunction{epa}{EPA}
\SetKwFunction{add}{add}
\KwIn{Hypergraph $G=(V,E)$}
\KwOut{The $(k,g)$-coreness of $G$}
$D \leftarrow \{\}$, 
$g \leftarrow 1$\;
\While{$true$}{
    $k = 0, H \leftarrow V, C \leftarrow \{\}, VQ \leftarrow \queueInit()$\;
    $T \leftarrow$ \buildBuckets{}; \tcp{Store nodes grouped by $g$-neighbour size}
    \For{$v \in H$}{
        $N_g (v) \leftarrow$ \getNbrMap{$H, E, v, g$}\;
        $C[v] \leftarrow \left|N_g (v)\right|$\;
        \If{$T[C[v]] = \emptyset$}{
            $T[C[v]] = \{\}$\;
        }
        $T[C[v]].$\add{$v$} \tcp{Group nodes by $g$-neighbour size}
    }

    \If{$T$.\size() $= 0$}{
        \break\;
    }

    \While{$H \neq \emptyset$}{
        $k \leftarrow k + 1$\;

        \For{$j \in$ \getKeys{$T$}}{
            \If{$j \geq k$}{
                \textbf{break}
            }
            \Else{
                $VQ$.\pushAll{$T[j]$}\;
            }
        }

        \While{$VQ$.\emptyc{} $\neq$ \true}{
            $v \leftarrow VQ$.\pop{}\; 
            $N_g (v) \leftarrow$ \getNbrMap{$H, E, v, g$}\;
            $H \leftarrow H \setminus \{v\}$\; 
            \For{$w \in$ \getKeys{$N_g(v)$}}{
                \If{$VQ$.\contain{$w$} $\neq$ \true}{
                    $T[C[w]].$\remove{$w$}\;
                    $C[w] \leftarrow C[w]-1$\;
                    \If{$C[w]<k$}{
                        $VQ$.\push{$w$}\;
                    }
                    \If{$T[C[w]] = \emptyset$}{
                        $T[C[w]] = \{\}$\;
                    }
                $T[C[w]].$\add{$w$}\; 
                    
                }
            }
        }
        $D[(k,g)] \leftarrow H$\;
    }
    $g \leftarrow g+1$\;
}
$D$ $\leftarrow$ $\Deduplication{D}$\;
\Return{$D$}
\caption{\mbox{Bucket-based coreness algorithm (\BCA)}}
\label{alg:kgcore_decomposition}
\end{algorithm}

\begin{algorithm}[h]
\small
\SetAlgoLined
\SetKwData{false}{false}
\SetKwData{true}{true}
\SetKwData{del}{delete}
\SetKwData{continue}{continue}
\SetKwFunction{queueInit}{queue}
\SetKwFunction{emptyc}{isEmpty}
\SetKwFunction{exist}{exist}
\KwIn{A collection $\mathcal{D}$ of $(k, g)$-pairs mapped to nodes}
\KwOut{A redundancy-free collection $\mathcal{D'}$ of $(k, g)$-pairs mapped to nodes}
$D' \leftarrow \{\}$\; 
\For{($k,g$) $\in$ \getKeys{$D$}}{
    $V' \leftarrow D[(k,g)]$\;
    \If{\exist{$D[(k+1,g)]$}}{
        $V' \leftarrow V' \setminus D[(k+1,g)]$\;
    }
    \If{\exist{$D[(k,g+1)]$}}{
        $V' \leftarrow V' \setminus D[(k,g+1)]$\;
    }
    \If{$V'$.\emptyc{} $\neq$ $\true$}{
        $D'[(k,g)] \leftarrow V'$\;
    }
    \Else{
        \continue\;
    }
    
}

\Return{$D'$}
\caption{\FuncSty{Deduplication}: Eliminating redundant entries in $(k,g)$-core }
\label{alg:reorganise}
\end{algorithm}

\spara{Pseudo Description.} 
Algorithm~\ref{alg:kgcore_decomposition} describes the process of {\BCA}. Initially, $k$ is set to $0$ and $g$ to $1$, with all storage structures initialised (Lines 1-4). For each node, the number of $g$-neighbours is computed, and nodes are assigned to buckets based on their $g$-neighbour count (Lines 5-10). 
These buckets store nodes grouped according to their $g$-neighbour count, unlike {\EPA}, which stores the $g$-neighbour count for each node. This allows for more efficient peeling by allowing us to focus on only those nodes whose $g$-neighbour count falls below a $k$ threshold rather than checking all nodes.

Next, nodes that do not satisfy the $k$ constraint are marked for removal (Lines 14-20). These nodes are deleted from the subhypergraph, and the $g$-neighbours of the removed nodes are reassigned to the corresponding buckets after reducing their $g$-neighbour count by one (Lines 21-27, 32). Any node whose updated $g$-neighbour count falls below $k$ is also marked for removal and deleted in the next iteration (Line 28).  

This procedure is repeated for all possible values of $g$ (Line 34). Once $(k,g)$-cores have been computed for all values of $k$ and $g$, Algorithm~\ref{alg:reorganise} removes redundancies to compute the final $(k,g)$-coreness values. It checks if a node in the $(k,g)$-core also presents in the $(k+1,g)$-core or $(k,g+1)$-core (Lines 3-7). If the node appears in a higher core value, it is removed from the current core (Lines 5,7), and iteratively checks all the computed $(k,g)$-cores (Line 2).

\begin{table}[t]
\centering
\caption{Initial $g$-neighbour counts in Figure~\ref{fig:example}}
\vspace{-0.3cm}
\label{tab:g-neighbour}
\vspace{-0.1cm}
\begin{subtable}{0.1667\textwidth}
\footnotesize
\centering
\subcaption{$g=1$}
\label{tab:g_1}
\begin{tabular}{c||c}
\hline \hline
$\boldsymbol{k}$ & \textbf{Nodes}  \\
\hline \hline
$2$ & $x_{11}$  \\
\hline
$3$ & $x_2$  \\
\hline
$4$ & $x_1,x_3,x_4,x_{10}$  \\
\hline
$5$ & $x_6, x_7, x_9$  \\
\hline 
$6$ & $x_5, x_8$  \\
\hline \hline
\end{tabular}
\end{subtable}%
\begin{subtable}{0.1667\textwidth}
\footnotesize
\centering
\subcaption{$g=2$}
\label{tab:g_2}
\begin{tabular}{c||c}
\hline \hline
$\boldsymbol{k}$ & \textbf{Nodes}  \\
\hline \hline
$0$ & $x_2, x_5, x_{10},x_{11}$  \\
\hline
$1$ & $x_9$  \\
\hline
$2$ & $x_1,x_3,x_4,x_6,x_7$  \\
\hline
$3$ & $x_8$  \\
\hline \hline
\end{tabular}
\end{subtable}%
\begin{subtable}{0.1667\textwidth}
\footnotesize
\centering
\subcaption{$g=3$}
\label{tab:g_3}
\begin{tabular}{c||c}
\hline \hline
$\boldsymbol{k}$ & \textbf{Nodes}  \\
\hline \hline
$0$ & $x_1, \cdots ,x_{11}$  \\
\hline \hline
\end{tabular}
\end{subtable}%
\end{table}

\begin{example}
Given the hypergraph in Figure~\ref{fig:example}, Table~\ref{tab:g-neighbour} shows the initial $g$-neighbour counts for different $g$ values. When the algorithm runs with $g=2$ and $k=1$, nodes $x_2, x_5, x_{10}$, and $x_{11}$, which have fewer than one $g$-neighbour, are removed. The number of $g$-neighbours for $x_8$, which is a $g$-neighbour of the removed nodes, is then updated. The bucket for nodes with two $g$-neighbours now contains $x_1, x_3, x_4, x_6, x_7$, and $x_8$.  

This process is repeated for increasing values of $k$ and $g$, producing all possible $(k,g)$-cores, as detailed in Table~\ref{tab:cores}. Finally, Algorithm~\ref{alg:reorganise} removes overlapping $(k,g)$-cores to obtain the final $(k,g)$-coreness values in Table~\ref{tab:coreness}.
\end{example}

\spara{Time Complexity.} The overall time complexity of {\BCA} is $O(g^*\cdot|e^*|\cdot \mathcal{D})$, where $g^*$ denotes the maximum $g$ value. Computing the $(k,g)$-coreness for all nodes requires $O(g^*\cdot|e^*|\cdot \mathcal{D})$ operations, as the process iterates up to $g^*$ times, with each iteration involving at most $|e^*|\cdot \mathcal{D}$ computations. In the step of eliminating redundant entries, duplicate nodes are removed through at most $k^*g^*$ set operations, each with a complexity of $O(|V|)$. As this complexity is significantly less intensive than that of the enumeration phase, the overall time complexity of {\BCA} remains $O(g^*\cdot|e^*|\cdot \mathcal{D})$.

\spara{Space Complexity.} 
The space complexity of {\BCA} is determined by three factors: storing the $(k,g)$-coreness of each node, maintaining the count of $g$-neighbours, and grouping nodes based on their $g$-neighbour count. Each node can have at most $\min(k^*, g^*)$ $(k,g)$-coreness values. Thus, the worst-case storage requirement of the $(k,g)$-coreness for every nodes is $O(\min(k^*, g^*) |V|)$.  During the peeling process, storing the number of $g$-neighbours for each node requires $O(|V|)$ space. Additionally, in each iteration, the structure that groups nodes based on $g$-neighbour count can store up to $O(|V|)$ nodes. Therefore, the overall space complexity remains $O(\min(k^*, g^*)|V|)$.

%% file: 05_experiments.tex
\section{EXPERIMENTS}
\label{sec:experiments}

In this section, we conduct comprehensive evaluations of the proposed Efficient Peeling Algorithm (\EPA), and Bucket-based Coreness Algorithm(\BCA). Through experiments, we observe the resultant subgraphs, running time and memory usage of the proposed algorithms on real-world and synthetic hypergraph datasets.
\subsection{Experimental Setup}
To evaluate {\EPA} and {\BCA}, we designed several experimental questions (EQs). The experimental questions are as follows:
\begin{itemize}[leftmargin=*]
    \item \textbf{EQ1. How do changes in the parameters $k$ and $g$ affect the cohesion of the resulting subhypergraphs?}
    This \textbf{EQ} examines the effects of varying parameters $k$ and $g$ on the cohesion of the resulting subhypergraphs, focusing on basic statistics such as the number of nodes, the number of hyperedges, average degree, and average support. 
    \item \textbf{EQ2. How does the running time of {\EPA} depend on the parameters?} This \textbf{EQ} analyses the running time of {\EPA} on various real-world datasets to identify performance trends and investigate the impact of varying $k$ and $g$ on computational efficiency.
    \item \textbf{EQ3. How scalable is {\EPA} as the dataset size increases?} This \textbf{EQ} evaluates the scalability of {\EPA} by analysing both the runtime performance and memory usage with varying the dataset sizes.
    \item \textbf{EQ4. How does {\EPA} consume memory throughout the computation process?} This \textbf{EQ} assesses the memory usage of {\EPA} throughout the entire computation process across real-world datasets.
    \item \textbf{EQ5. How does the $(k,g)$-core model compare with other models?} This \textbf{EQ} evaluates the $(k,g)$-core compared to other models, including clique-core~\cite{batagelj2011fast}, $(\alpha,\beta)$-core~\cite{alphabeta}, $k$-hypercore~\cite{leng2013m},$(k,t)$-hypercore~\cite{bu2023hypercore}, nbr-$k$-core~\cite{nbrkcore}, and $(k,d)$-core~\cite{nbrkcore}. 
    \item \textbf{EQ6. How does {\BCA} perform on running time?} This \textbf{EQ} evaluates the efficiency of {\BCA} by measuring its running time on both real-world and synthetic hypergraph datasets. 
    \item \textbf{EQ7. How are coreness values distributed in real-world dataset?}
    This \textbf{EQ} analyses the distribution of coreness values derived from two decomposition models, $(k,g)$-core and $(k,d)$-core, on the Meetup dataset.
    \item \textbf{EQ8. How effective is the $(k,g)$-core model in identifying frequent itemsets in case study?}  This \textbf{EQ} evaluates the ability of the $(k,g)$-core model to discover frequent itemsets through practical case study.
\end{itemize}

\subsection{Experimental Setting} \label{sec:exp_setting}
\begin{table*}[t]
\caption{Real-world datasets}
\vspace{-0.2cm}
\label{tab:data}
\centering

\begin{tabular}{c||c|c|c|c|c|c|c|c}
\hline \hline
\textbf{Dataset} & $\boldsymbol{|V|}$ & $\boldsymbol{|E|}$ &$\mathcal{D}$
&$\boldsymbol{k^*}$
&$\boldsymbol{g^*}$& $\boldsymbol{|N(.)|}$ & $\boldsymbol{|C(.)|}$ & $\boldsymbol{|D(.)|}$  \\ \hline \hline
Contact & $242$ & $12,704$ &$30,729$& $47$ & $54$ & $68.74$ & $2.42$ & $126.97$  \\ \hline
Congress & $1,718$ & $83,105$ &$732,300$ &$368$ & $1,003$ & $494.68$ & $8.81$ & $426.25$  \\ \hline
Meetup & $24,115$ & $11,027$ &$113,588$ &$121$& $250$ & $65.27$ & $10.3$ & $4.71$ \\ \hline
Kosarak & $41,270$ & $990,002$ & $8,018,988$ &$3,313$& $324,013$ & $1,604.18$ & $8.10$ &  $194.30$  \\ \hline
Walmart & $88,860$ & $69,906$ &$460,630$ &$127$& $729$ & $46.77$ & $6.59$ & $5.18$ \\ \hline
Gowalla & $107,092$ & $1,280,969$ & $3,981,334$ &$2,630$& $599$ & $916.20$ & $3.11$ & $37.17$ \\ \hline
Amazon & $2,268,231$ & $4,285,363$ & $73,141,425$ &$9,349$& $17,123$ & $2,673.81$ & $17.07$ & $32.24$ \\ \hline
AMiner & $27,850,748$ & $17,120,546$ & $64,555,978$ &$610$& $730$ & $8.39$ & $3.77$ & $2.31$ \\ \hline
\hline
\end{tabular}
\end{table*}

\spara{Parameters.}
Selecting a proper parameter value is challenging, even for $k$-core, which consists of only a single parameter~\cite{chu2020finding}. Determining universally applicable $(k,g)$-core parameters for all datasets is even more difficult. Thus, in our experiment, we set the default value as $k=5$ and $g=5$. To analyse the impact of parameter variations, we also explore $k$ and $g$ values of $3,4,5,6$, and $7$ in selected experiments and observe the resultant subgraphs. 

\spara{Datasets.}
We conduct experiments on a variety of real-world hypergraph datasets including different domains. These include a communication network (Contact~\cite{nbrkcore}), 
a co-sponsored bill network (Congress~\cite{nbrkcore}), social networks (Meetup~\cite{nbrkcore} and Gowalla~\cite{cho2011friendship}), 
co-authorship networks (AMiner~\cite{nbrkcore}), a clickstream network (Kosarak~\cite{Benson-2018-subset}), 
and co-purchase networks (Walmart~\cite{Amburg2020categorical} and Amazon~\cite{ni2019justifying}). 
These datasets vary in size and structural properties, allowing for a comprehensive evaluation of the proposed algorithm. 
Table~\ref{tab:data} summarises the key statistics of these datasets. These include $k^*$ and $g^*$ denote the maximum values of the parameters $k$ and $g$; and $|N(.)|$, $|C(.)|$, and $|D(.)|$ represent the average neighbour size, the average hyperedge cardinality, and the average degree, respectively.

\begin{table}[h]
\centering
\small
\caption{Parameter settings for each cohesive subgraph model.}
\vspace{-0.2cm}
\label{tab:model_parameters}
\begin{tabular}{c||c|c|c}
\hline \hline
\textbf{Model} & \multicolumn{2}{c|}{\textbf{Constraints}} & \textbf{Type} \\ \hline \hline
$k$-hypercore~\cite{leng2013m} & \multicolumn{2}{c|}{Degree \footnotesize{$(k=5)$}} & Hypergraph \\ \hline
nbr-$k$-core~\cite{nbrkcore} & \multicolumn{2}{c|}{Neighbour size \footnotesize{$(k=5)$}} & Hypergraph \\ \hline
\multirow{2}{*}{$(k,t)$-hypercore~\cite{bu2023hypercore}}
    & Degree & Fraction & \multirow{2}{*}{Hypergraph} \\
    & \footnotesize{$(k=5)$} & \footnotesize{$(t=0.5)$} &  \\ \hline
\multirow{2}{*}{$(k,d)$-core~\cite{nbrkcore}}
    & Neighbour size & Degree & \multirow{2}{*}{Hypergraph} \\
    & \footnotesize{$(k=5)$} & \footnotesize{$(d=5)$} &  \\ \hline
\multirow{2}{*}{$(\alpha,\beta)$-core~\cite{alphabeta}} 
    & Degree & Cardinality & \multirow{2}{*}{Bipartite} \\
    & \footnotesize{$(\alpha=5)$}& \footnotesize{$(\beta=5)$} &  \\ \hline
clique-core~\cite{batagelj2011fast}
    & \multicolumn{2}{c|}{Neighbour size \footnotesize{$(k=5)$}} & Unipartite \\ \hline \hline
\end{tabular}
\vspace{-0.4cm}
\end{table}

\spara{Algorithm.} We compare six existing cohesive subgraph models as baselines, including $k$-hypercore, nbr-$k$-core, $(k,d)$-core, $(k,t)$-hypercore, $(\alpha,\beta)$-core, and clique-core. Note that $k$-hypercore, nbr-$k$-core, and $(k,d)$-core~\cite{nbrkcore} identify strongly induced subgraphs, which means that any node removed must also discard all hyperedges containing it.
The $(\alpha,\beta)$-core and clique-core~\cite{batagelj2011fast} apply by converting hypergraphs into bipartite and unipartite subgraphs, respectively. The constraints and parameter settings for each model are summarised in Table~\ref{tab:model_parameters}.

\subsection{Experimental Results}

\spara{EQ1. Impact of Parameter Variations on Subhypergraph Cohesion.}  
We examine the trends in how parameter variations affect subhypergraph structures across four main statistics: the number of nodes, the number of hyperedges, average degree, and average support.  
To observe the impact of changing $g$ while keeping $k$ constant, we fix $k$ as the default parameter and vary $g$ over the values $3,4,5,6$, and $7$.  Similarly, to analyse the effects of changing $k$ values with a constant $g$, we fix $g$ as the default parameter and vary $k$ over the values $3,4,5,6$, and $7$.

\begin{figure}[h]
    \begin{subfigure}{.99\linewidth} 
        \centering
        \includegraphics[width=0.99\linewidth]{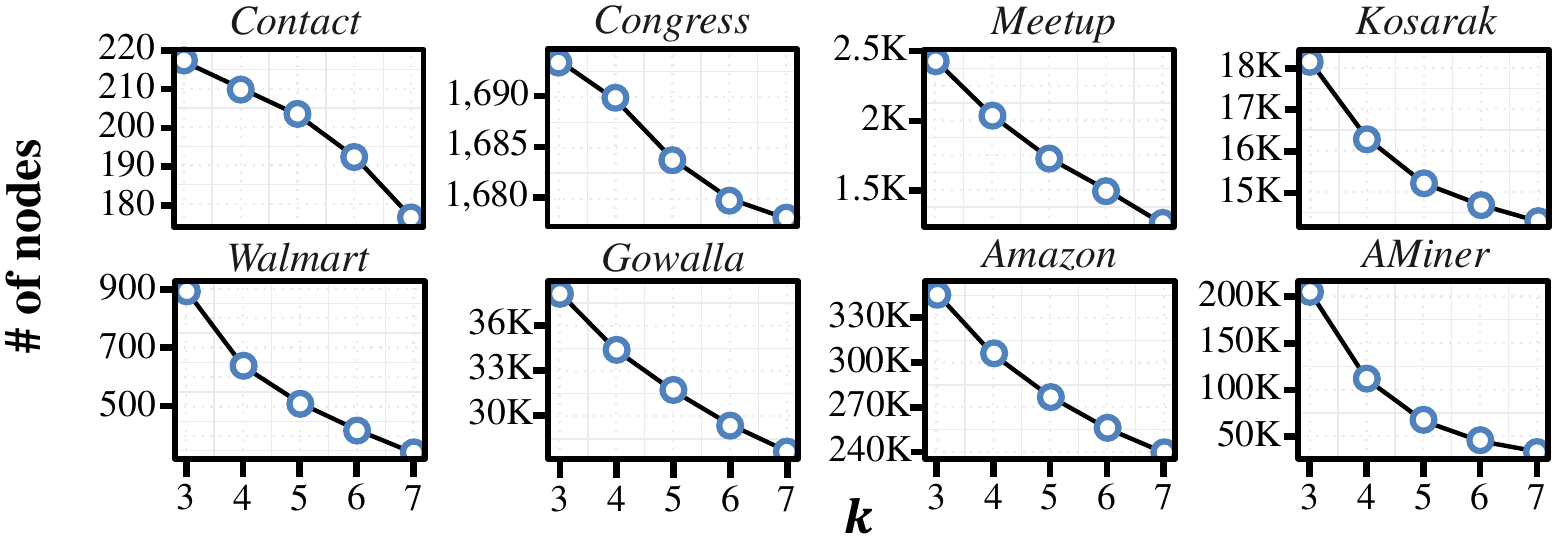}
        \vspace{-0.2cm}
        \caption{Varying of the parameter $k$}
        \vspace{0.1cm}
        \label{fig:eq1_1_k}
    \end{subfigure}
    \begin{subfigure}{.99\linewidth} 
        \centering
        \includegraphics[width=0.99\linewidth]{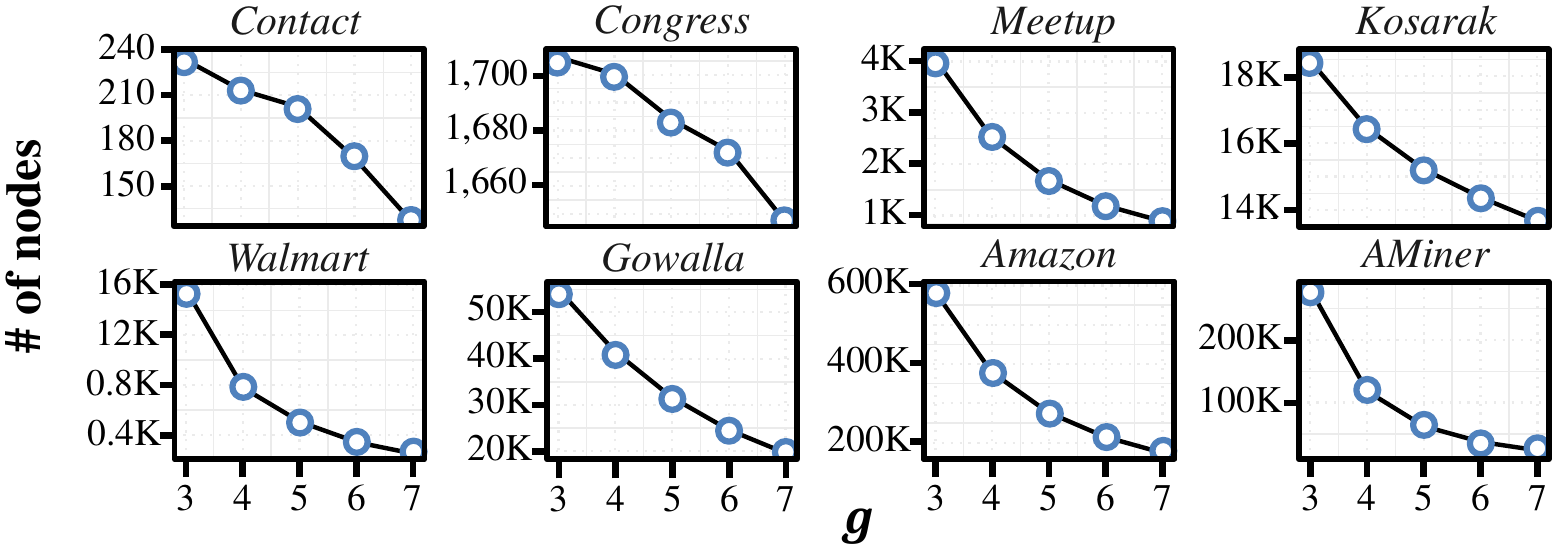}
        \vspace{-0.2cm}
         \caption{Varying of the parameter $g$}
        \label{fig:eq1_1_g}
    \end{subfigure}    
    \vspace{-0.3cm}
    \caption{EQ1-1. Impact of parameter variations on node size}    
    \vspace{-0.3cm}
    \label{fig:eq1_1}
\end{figure}

\spara{EQ1-1. Impact of Parameter Variations on Subhypergraph Cohesion (Number of Nodes).}  
Figure~\ref{fig:eq1_1} illustrates the effect of parameter variations on node count.  
We observe that as $k$ increases, the size of the subhypergraph decreases.  
This occurs because stricter neighbour constraints reduce the number of nodes that satisfy the conditions, as more neighbours are required for inclusion.  
Similarly, increasing $g$ leads to a smaller cohesive subgraph, as a higher co-occurrence threshold further restricts node inclusion, resulting in fewer nodes meeting the constraint.

\begin{figure}[h]
    \begin{subfigure}{.99\linewidth} 
        \centering
        \includegraphics[width=0.99\linewidth]{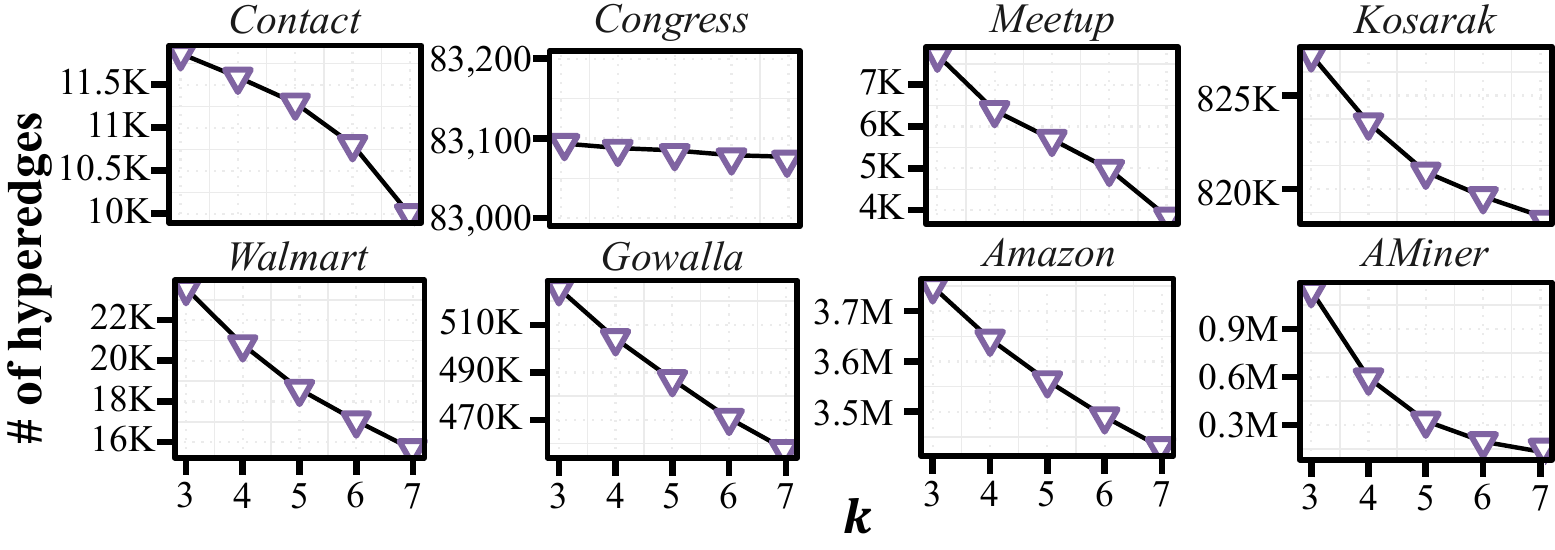}
        \vspace{-0.2cm}
        \caption{Varying of the parameter $k$}
        \vspace{0.1cm}
        \label{fig:eq1_4_k}
    \end{subfigure}
    \begin{subfigure}{.99\linewidth} 
        \centering
        \includegraphics[width=0.99\linewidth]{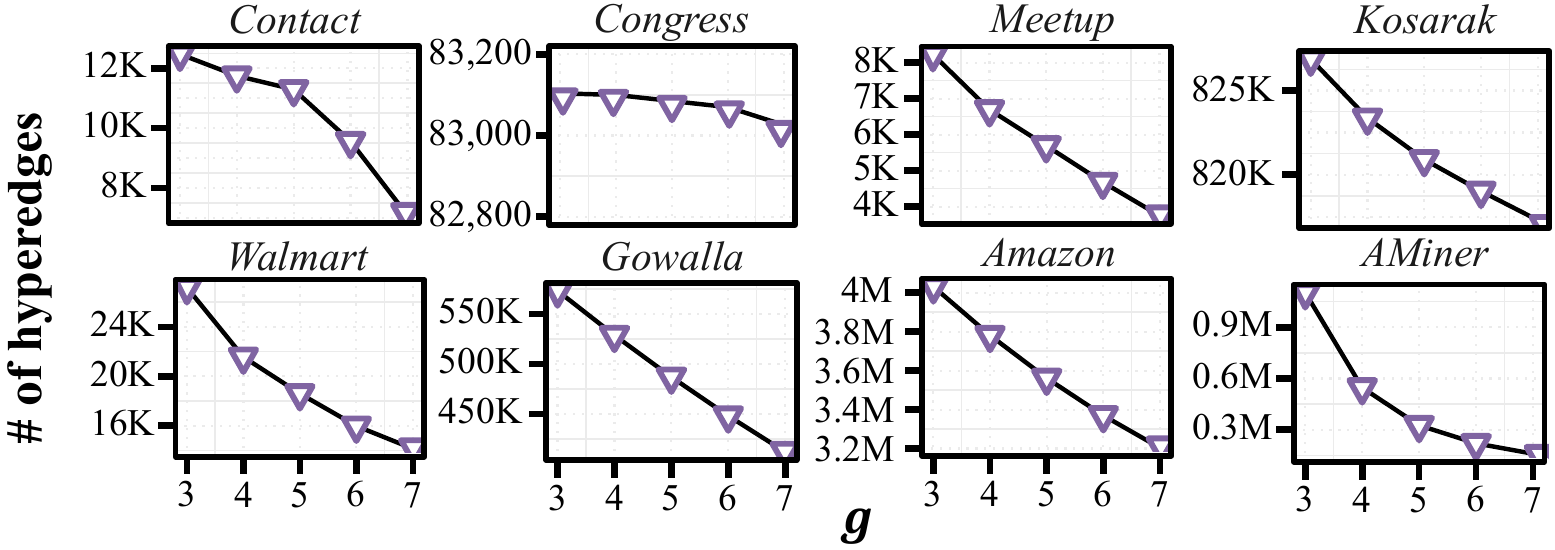}
         \vspace{-0.2cm}
         \caption{Varying of the parameter $g$}
        \label{fig:eq1_4_g}
    \end{subfigure}    
    \vspace{-0.3cm}
    \caption{EQ1-2. Impact of parameter variations on the number of hyperedges}    
    \vspace{-0.4cm}

    \label{fig:eq1_4}
\end{figure}

\spara{EQ1-2. Impact of Parameter Variations on Subhypergraph Cohesion (Number of Hyperedges).} The effect of $k$ and $g$ variations on the number of hyperedges can be observed in Figure~\ref{fig:eq1_4}. As $k$ increases, the reduction in node count due to stricter neighbour constraints also leads to a decrease in the number of hyperedges. This occurs because fewer nodes satisfy the condition, and as a result, the hyperedges that contain those nodes also decrease in number. Similarly, when $g$ increases, the higher co-occurrence threshold reduces the number of nodes that can be included, thus further decreasing the number of hyperedges. In the case of the Congress dataset, the same trend is observed, but the extend of the decrease is smaller compared to the other datasets. This is because when the parameters are varied, fewer nodes are removed compared to other datasets (refer to Figure~\ref{fig:eq1_1}), which results in more hyperedges remaining. Even though a few nodes may be removed from a specific hyperedge, other nodes in that hyperedge still satisfy the constraints, allowing the hyperedge to be preserved. These results suggest that the Congress dataset can be considered a network with strong connectivity, where the structure is well-maintained even under stricter constraints.

\begin{figure}[h]
    \begin{subfigure}{.99\linewidth} 
        \centering
        \includegraphics[width=0.99\linewidth]{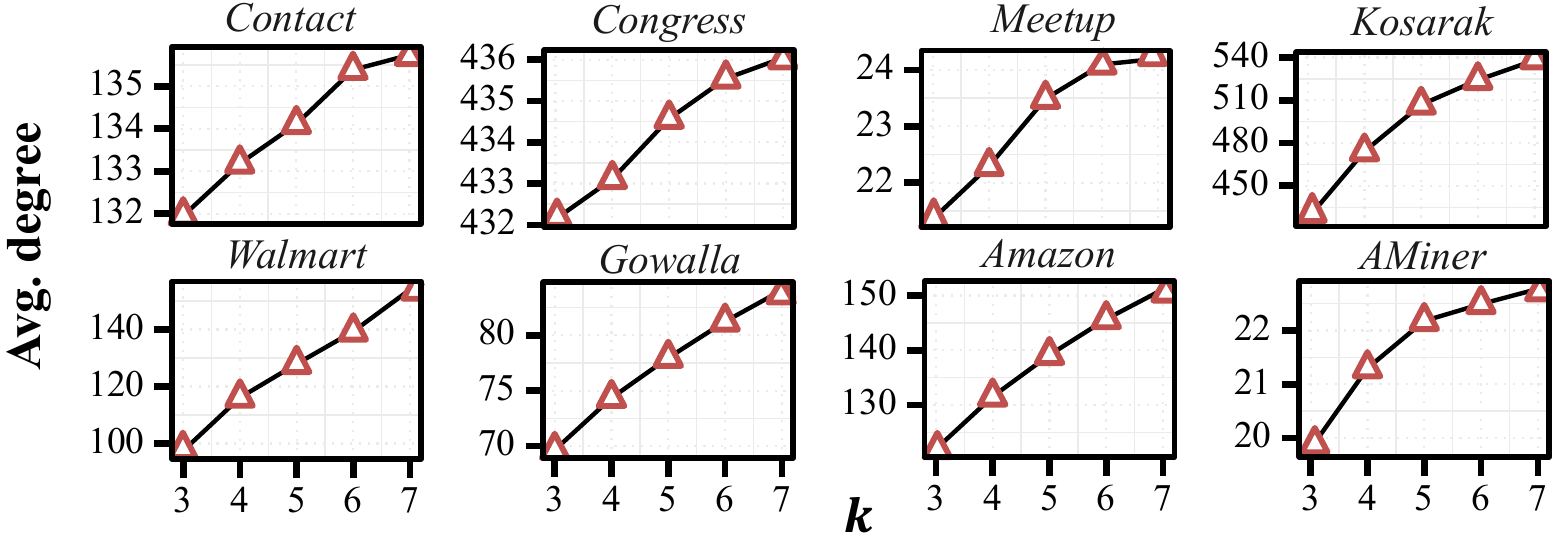}   
        \vspace{-0.2cm}
        \caption{Varying of the parameter $k$}
        \vspace{0.1cm}
        \label{fig:eq1_2_k}
    \end{subfigure}
    \begin{subfigure}{.99\linewidth} 
        \centering
        \includegraphics[width=0.99\linewidth]{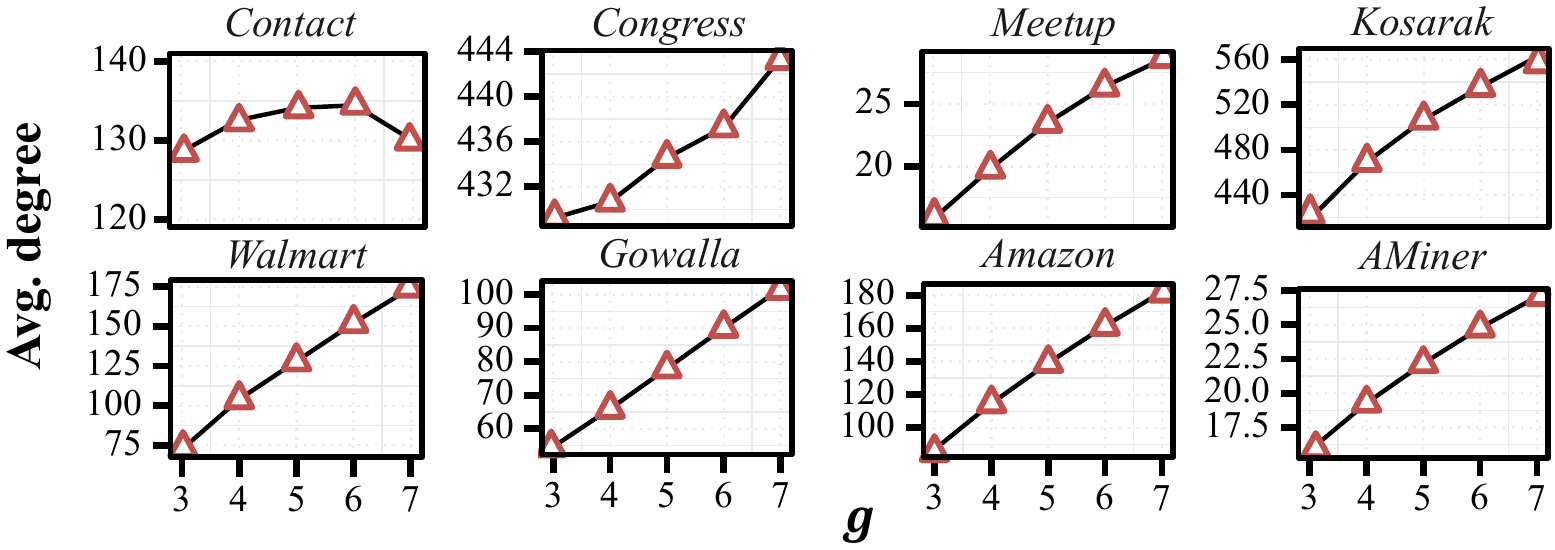}
        \vspace{-0.2cm}
        \caption{Varying of the parameter $g$}
        \label{fig:eq1_2_g}
    \end{subfigure}
    \vspace{-0.3cm}
    \caption{EQ1-3. Impact of parameter variations on average degree}
    \vspace{-0.3cm}
    \label{fig:eq1_2}
\end{figure}

\spara{EQ1-3. Impact of Parameter Variations on Subhypergraph Cohesion (Average Degree).}  
We analyse how changes in $k$ and $g$ affect the average degree of the subhypergraph, where the average degree represents the average number of hyperedges each node belongs to. Figures~\ref{fig:eq1_2_k} and \ref{fig:eq1_2_g} respectively illustrate how the average degree changes when $k$ increases with $g$ set to its default value, and when $g$ increases with $k$ set to its default value. Overall, we observe that increasing $k$ or $g$ leads to a rise in the average degree. As $k$ increases, nodes with fewer connections are removed, resulting in a more densely connected subhypergraph with a higher average degree. Similarly, as $g$ increases, nodes that participate in fewer hyperedges are removed, which leads to a higher average degree. 
However, in the Contact dataset, the average degree slightly drops when $g$ reaches $7$. This case occurs due to a substantial reduction in hyperedges resulting from stricter co-occurrence constraints. Specifically, compared to the configuration $(k=5,g=6)$, the number of hyperedges is reduced by about $19.45$\%, which is significantly higher than the average reduction of $4.31\%$.

\begin{figure}[h]
    \begin{subfigure}{.99\linewidth} 
        \centering
        \includegraphics[width=0.99\linewidth]{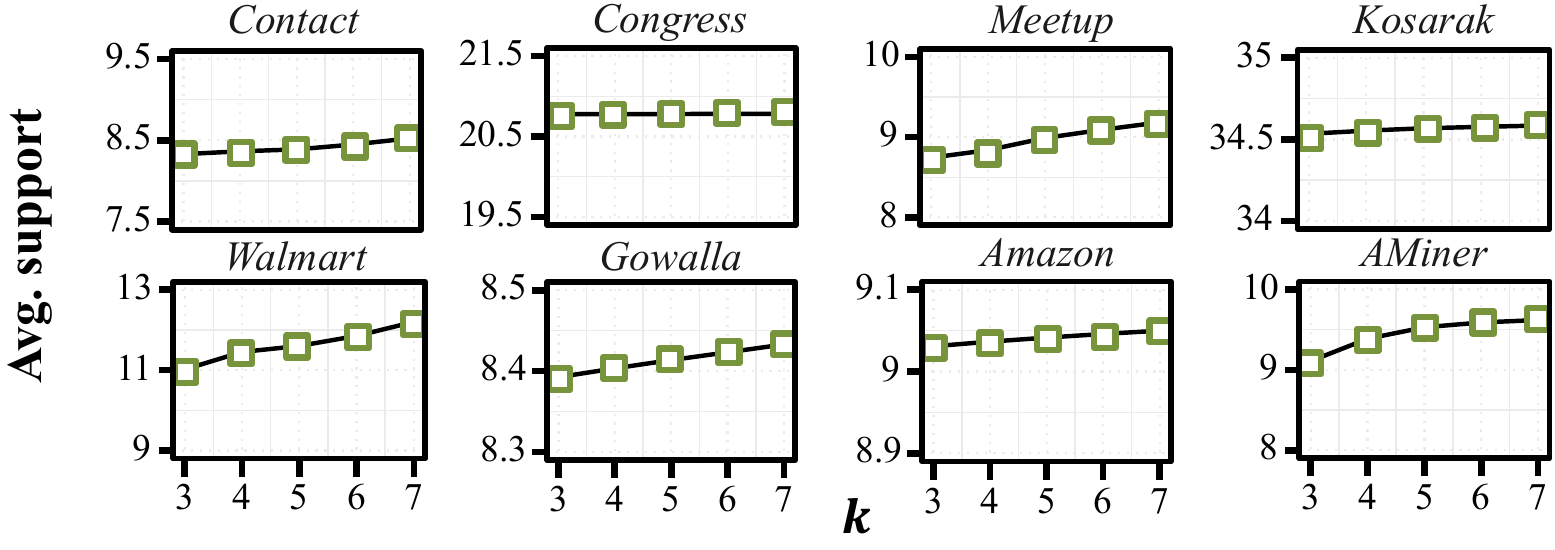}         
        \vspace{-0.2cm}
        \caption{Varying of the parameter $k$}
         \vspace{0.1cm}
        \label{fig:eq1_3_k}
    \end{subfigure}
    \begin{subfigure}{.99\linewidth} 
        \centering
        \includegraphics[width=0.99\linewidth]{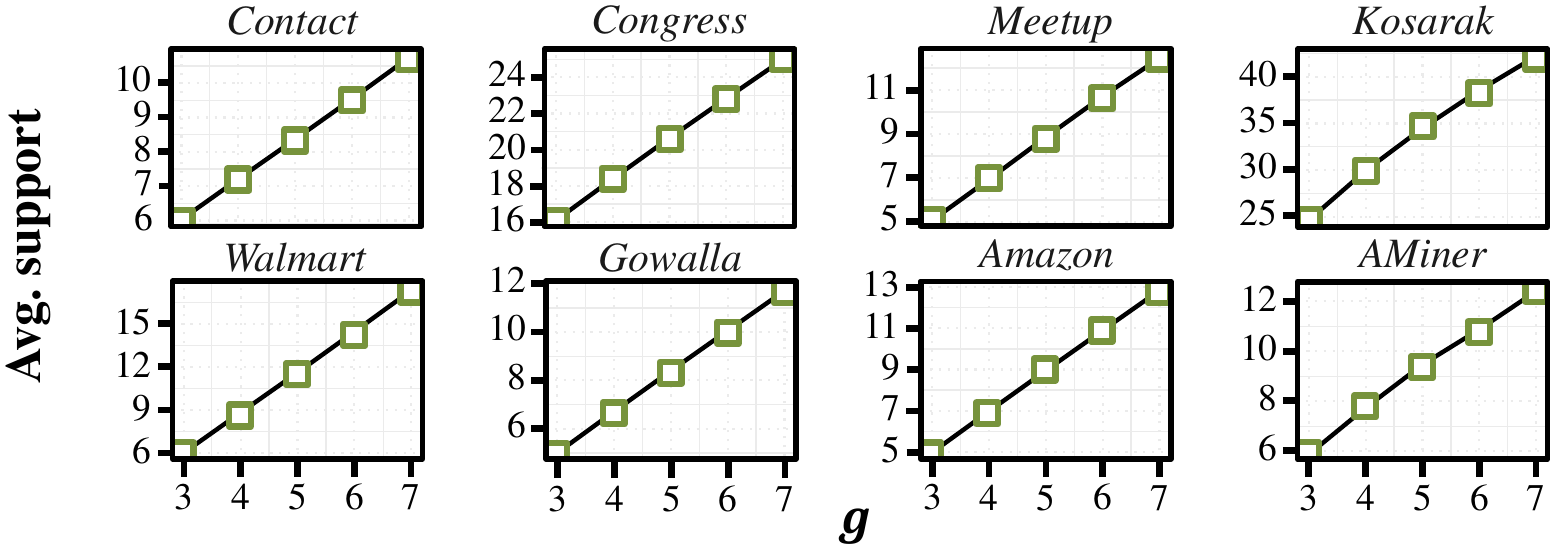}        
        \vspace{-0.2cm}
        \caption{Varying of the parameter $g$}
        \label{fig:eq1_3_g}
    \end{subfigure}
    \vspace{-0.3cm}
    \caption{EQ1-4. Impact of parameter variations on average support}
    \vspace{-0.5cm}
    \label{fig:eq1_3}
\end{figure}

\spara{EQ1-4. Impact of Parameter Variations on Subhypergraph Cohesion (Average Support).}  
The average support is defined as the average support value across all neighbour node pairs in the resulting subhypergraph. We analyse the impact of average support as parameter are varied.  
Figure~\ref{fig:eq1_3_k} shows the average support as $k$ increases.  
We observe that the average support slightly increases  with $k$.  
This is because stricter neighbour constraints result in the removal of less cohesive nodes from the resulting subhypergraph.
As shown in Figure~\ref{fig:eq1_3_g}, the average support also increases as $g$ increases, and the increase is relatively large compared to the extent of the increase with changes in $k$ across all datasets.
A stricter co-occurrence threshold allows only node pairs with high co-occurrence frequency to be included in the subhypergraph.  
As a result, nodes with fewer co-occurrences are removed, leading to a notable increase in the average support.  
Notably, Kosarak exhibits a higher average support than other datasets, likely reflecting its relatively high average degree and, consequently, more frequent co-occurrences among nodes.

\begin{figure}[!b]
    \begin{subfigure}{.99\linewidth} 
        \centering
        \includegraphics[width=0.99\linewidth]{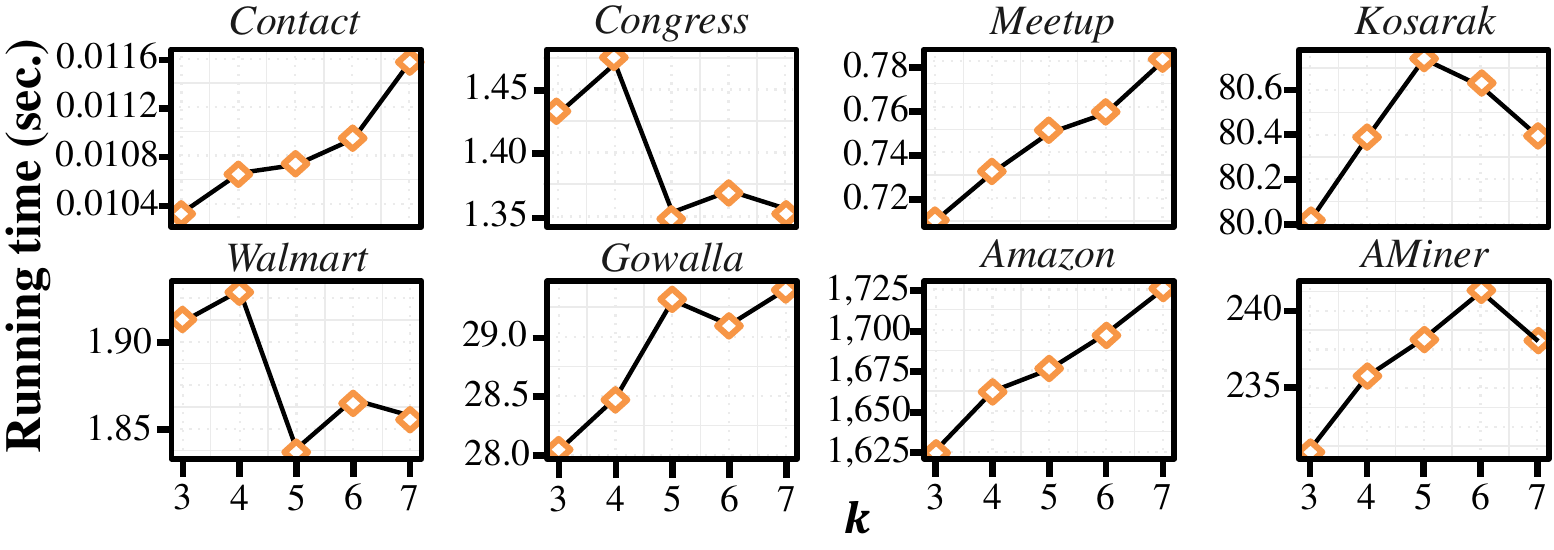}
        \vspace{-0.2cm}
        \caption{Varying parameter $k$}
        \vspace{0.2cm}
        \label{fig:eq2_g}
    \end{subfigure}
    \hfill
    \begin{subfigure}{.99\linewidth} 
        \centering
        \includegraphics[width=0.99\linewidth]{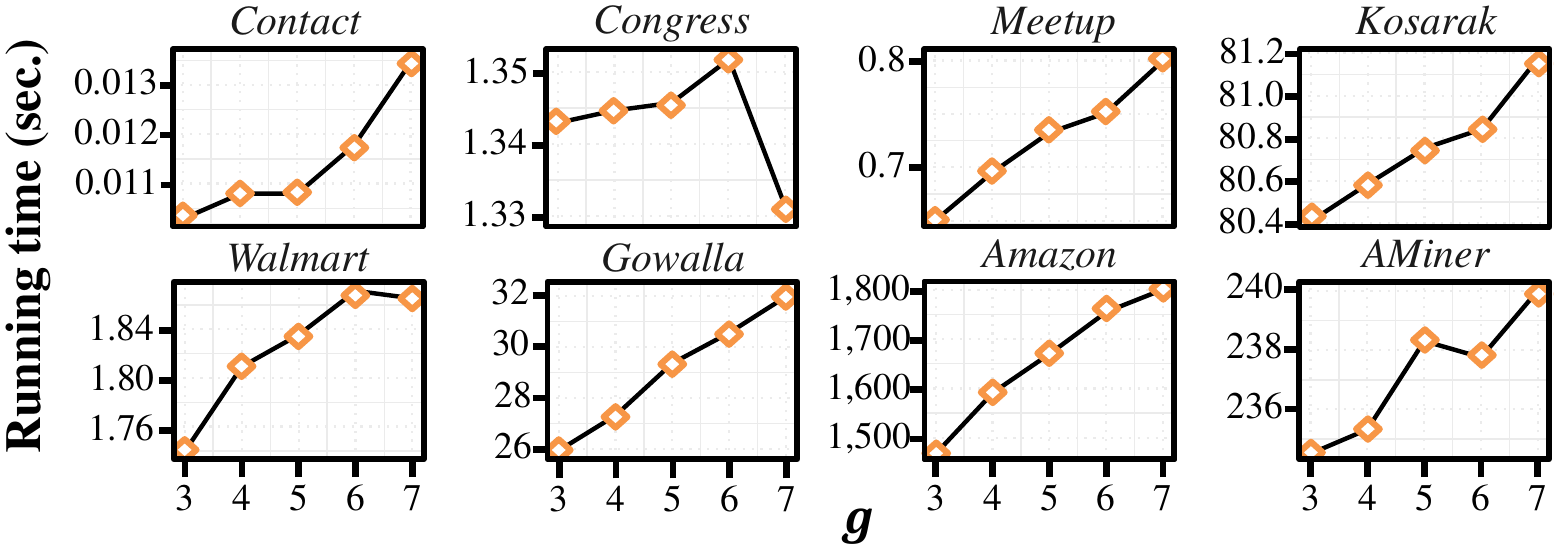}
        \vspace{-0.2cm}
        \caption{Varying parameter $g$}
        \label{fig:eq2_k}
    \end{subfigure}
    \vspace{-0.3cm}
    \caption{EQ2. Running time with varying parameters}
    \vspace{-0.4cm}
    \label{fig:eq2}
\end{figure}
\spara{EQ2. Running Time with Varying Parameters.} 
This experiment evaluates how changes in the parameters $k$ and $g$ impact running time across datasets of varying sizes and structural properties.  
For larger datasets such as Amazon and AMiner, where the number of nodes and hyperedges is substantial, the execution time is significantly longer (see Figure~\ref{fig:eq2}).  
Additionally, in datasets with a large average neighbour size, such as Amazon and Kosarak, the runtime further increases as each node must verify a greater number of $g$-neighbours during the peeling process.  
We also analyse the effect of varying $k$ and $g$ within each dataset.  
As illustrated in Figures~\ref{fig:eq2_g} and~\ref{fig:eq2_k}, increasing $k$ or $g$ imposes stricter constraints on the $(k,g)$-core, leading to the cascading removal of more nodes and their associated $g$-neighbours, thereby extending the execution time, particularly in datasets with a high number of hyperedges.  
Conversely, for smaller datasets such as Contact, Congress, Meetup, and Walmart, the execution time remains relatively stable, with variations of less than a second across all parameter settings.

\begin{figure}[h]
    \centering
    \begin{subfigure}[b]{0.49\linewidth} 
        \centering
        \includegraphics[width=\linewidth]{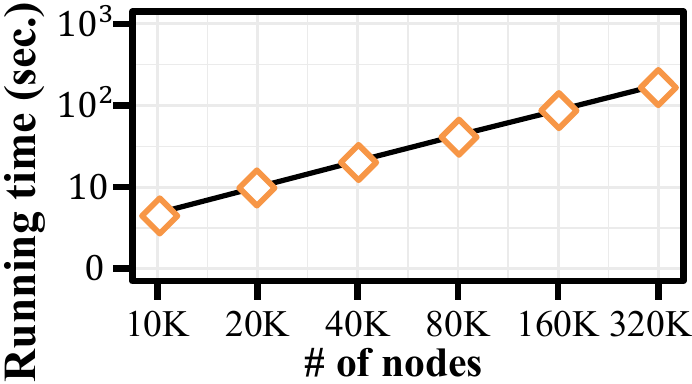}
        \vspace{-0.5cm}
        \caption{Running time}
        \label{fig:eq4_1}
    \end{subfigure}
    \begin{subfigure}[b]{0.49\linewidth} 
        \centering
        \includegraphics[width=\linewidth]{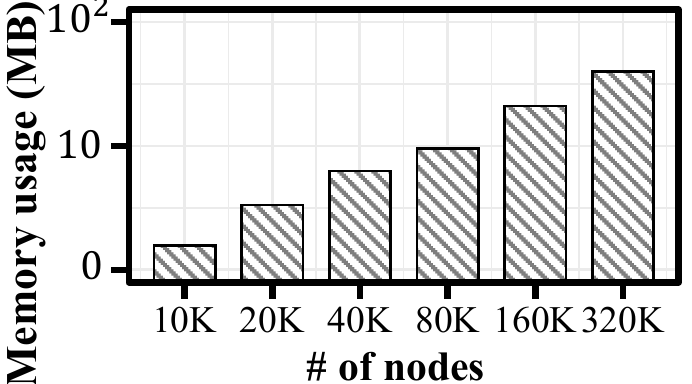}
        \vspace{-0.5cm}
        \caption{Memory usage}
        \label{fig:eq4_2}
    \end{subfigure}
    \vspace{-0.2cm}
    \caption{EQ3. Scalability test}
    \vspace{-0.2cm}
    \label{fig:scalability}
\end{figure}

\spara{EQ3. Scalability.}
Figure~\ref{fig:scalability} shows the scalability of {\EPA} by demonstrating how running time and memory consumption evolve as the hypergraph grows from $10,000$ to $320,000$ nodes.  
We generated these synthetic hypergraphs using the H-ABCD benchmark model~\cite{kaminski2023hypergraph}, with a node degree distribution of $2.1$ (ranging from $40$ to $1,000$) and a community size distribution of $1.7$ (from $80$ to $900$ communities).  
Hyperedge sizes were selected uniformly between $1$ and $40$, with a noise factor of $0.2$.
The experiment was conducted using the default parameters. Figure~\ref{fig:eq4_1} presents the running time in relation to the number of nodes in the hypergraph, revealing an almost linear trend, where execution time increases from $4.32$ seconds for $10,000$ nodes to $164.18$ seconds for $320,000$ nodes.  
Similarly, Figure~\ref{fig:eq4_2} illustrates peak memory usage, which also exhibits a nearly linear trend, growing from $1.57$MB at $10,000$ nodes to $40.07$MB at $320,000$ nodes.  
Overall, these findings indicate that the proposed algorithm scales efficiently in both time and memory.  

\begin{figure}[h]
    \centering
    \includegraphics[width=0.85\linewidth]{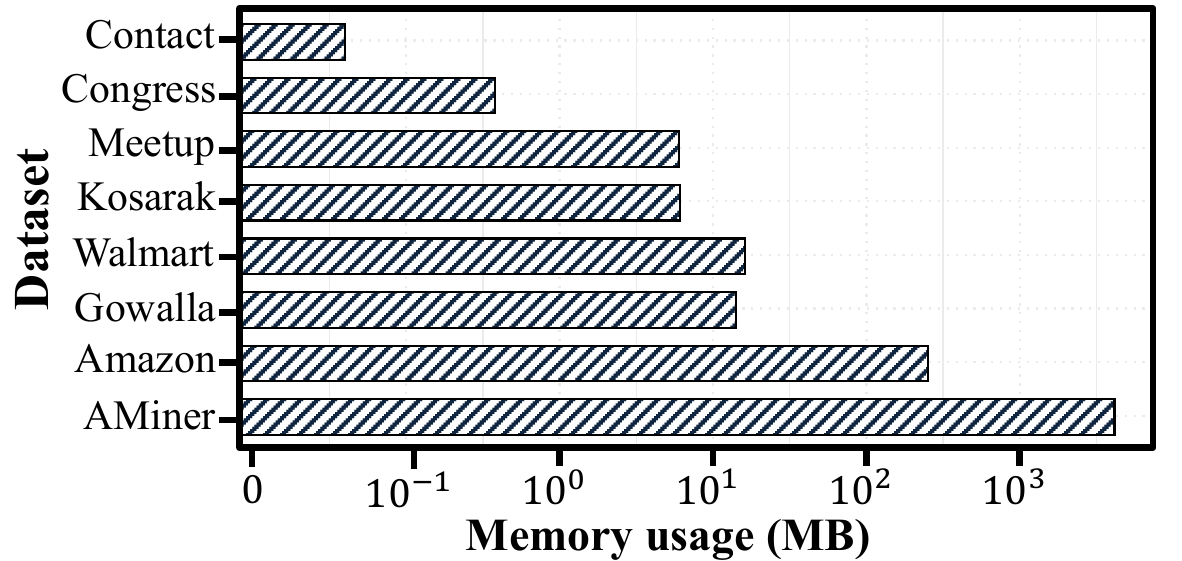}
    \vspace{-0.2cm}
    \caption{EQ4. Evaluation of memory usage}
    \vspace{-0.2cm}
    \label{fig:eq3}
 \end{figure}

\spara{EQ4. Evaluation of Memory Usage.}
As shown in Figure~\ref{fig:eq3}, we evaluate the peak memory consumption of {\EPA} across different datasets with default parameters. Memory usage, in general, tends to increase with the number of nodes, which requires storage of $g$-neighbour count. Large-scale datasets such as AMiner and Amazon exhibit higher peak memory usage. However, memory usage may vary with the structural characteristics of each dataset, including average hyperedge cardinality and average neighbour size. Walmart had higher memory usage than Gowalla, indicating that average hyperedge cardinality may affect memory consumption, especially in datasets with similar node sizes.

\spara{EQ5. Comparison of $(k,g)$-core with other models.} In this experiment, we compare the $(k,g)$-core model to six existing cohesive subgraph models, the description and the parameter settings of the models are presented in Section~\ref{sec:exp_setting}. We evaluate and compare the resultant subhypergraphs by examining three key statistics: vertex density, average degree, and average support. 

\begin{figure}[h]
    \centering
    \includegraphics[width=0.99\linewidth]{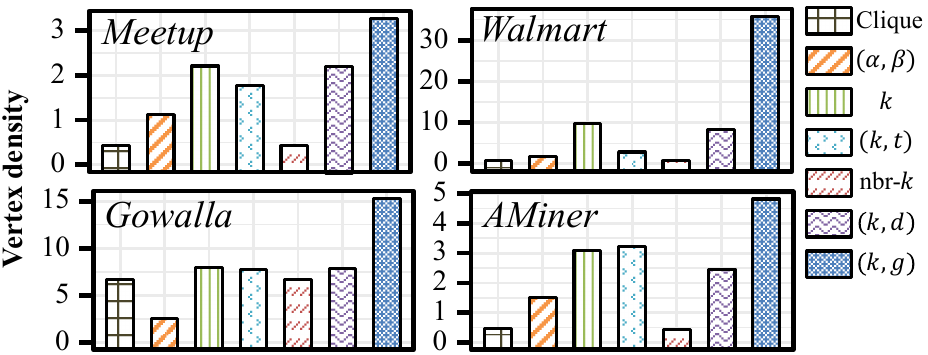}
    \vspace{-0.2cm}
    \caption{EQ5-1. Comparison of vertex density in the resultant subhypergraphs}
    \vspace{-0.2cm}
    \label{fig:eq7_density}
 \end{figure}

\spara{EQ5-1. Comparison of $(k,g)$-core with Other Models (Vertex density).} We compare vertex density—defined as the number of hyperedges divided by the number of nodes in the resulting subhypergraph—across various cohesive subhypergraph models, as shown in Figure~\ref{fig:eq7_density}. The $(k,g)$-core  shows consistently higher vertex density than  all baselines across datasets. In particular, on the Walmart dataset, the vertex density of the $(k,g)$-core is over $47.7$ times greater than that of the nbr-$k$-core and nearly $3.6$ times higher than the best-performing $k$-hypercore. These results demonstrate that by incorporating both degree and co-occurrence constraints, the $(k,g)$-core effectively filters out loosely connected nodes and retains groups with strong internal connectivity, resulting in dense subhypergraphs.

\begin{figure}[h]
    \centering
    \includegraphics[width=0.99\linewidth]{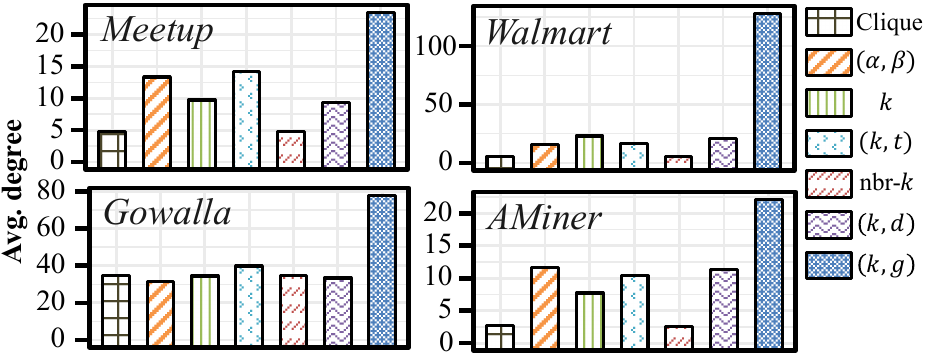}
    \vspace{-0.2cm}
    \caption{EQ5-2. Comparison of the average degree in the resultant subhypergraphs}
    \vspace{-0.2cm}
    \label{fig:eq7_deg}
 \end{figure}

\begin{figure}[h]
    \centering
    \includegraphics[width=0.99\linewidth]{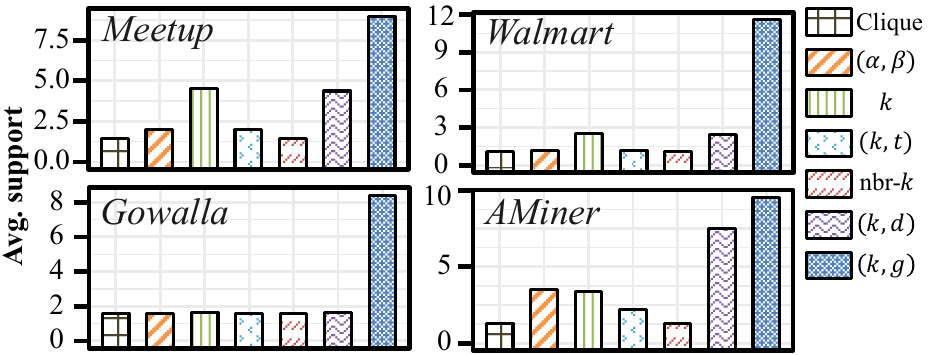}
    \vspace{-0.2cm}
    \caption{EQ5-3. Comparison of the average support in the resultant subhypergraphs}
    \vspace{-0.2cm}
    \label{fig:eq7_sup}
 \end{figure}
 
\spara{EQ5-2. Comparison of $(k,g)$-core with Other Models (Average Degree).} We analyse the average degree of nodes in the resulting subhypergraphs, which is represented in Figure~\ref{fig:eq7_deg}. The $(k,g)$-core yields notably higher average degrees compared to other models. This is because the parameter $g$ of $(k,g)$-core implicitly reflects a degree constraint: for a node to co-occur with at least $g$ other nodes, it must participate in at least $g$ hyperedges.  Consequently, nodes retained in the $(k,g)$-core tend to have frequent hyperedge participation, resulting in a higher average degree. Compared to the $(k,g)$-core, other models such as the $(\alpha,\beta)$-core, $k$-hypercore, $(k,t)$-hypercore, and $(k,d)$-core yield lower average degrees, reflecting their weaker cohesion constraints.

\begin{figure}[h]
    \centering
    \begin{subfigure}[h]{0.49\linewidth} 
        \centering
        \includegraphics[width=\linewidth]{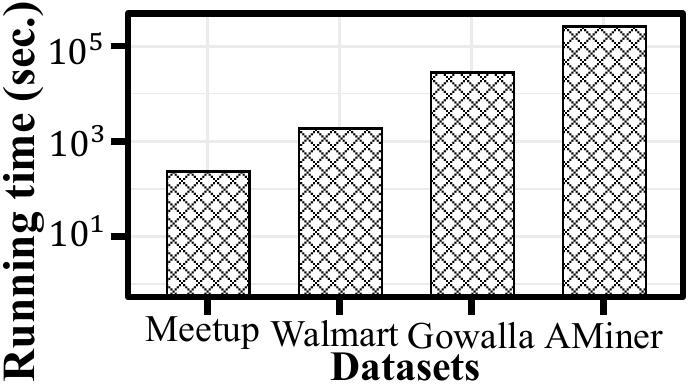}
        \vspace{-0.5cm}
        \caption{Real-world datasets}
        \label{fig:eq6_1}
    \end{subfigure}
    \begin{subfigure}[h]{0.49\linewidth} 
        \centering
        \includegraphics[width=\linewidth]{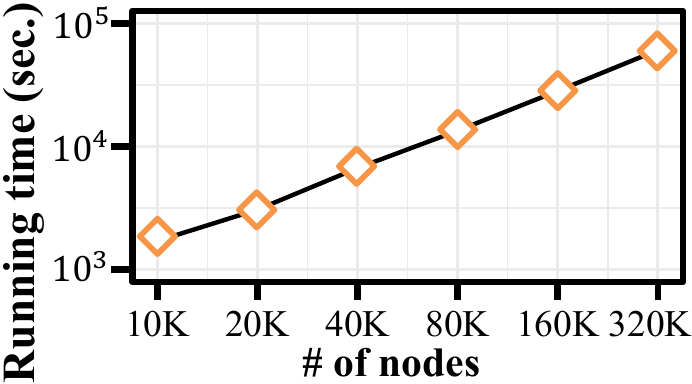}
        \vspace{-0.5cm}
        \caption{Synthetic datasets}
        \label{fig:eq6_2}
    \end{subfigure}
    \vspace{-0.2cm}
    \caption{EQ6. Running time of {\BCA}}
    \vspace{-0.5cm}
    \label{fig:eq6_bca}
\end{figure}

\begin{figure*}[!b]
    \centering
    \begin{subfigure}[h]{0.90\linewidth} 
        \centering
        \includegraphics[width=0.99\linewidth]{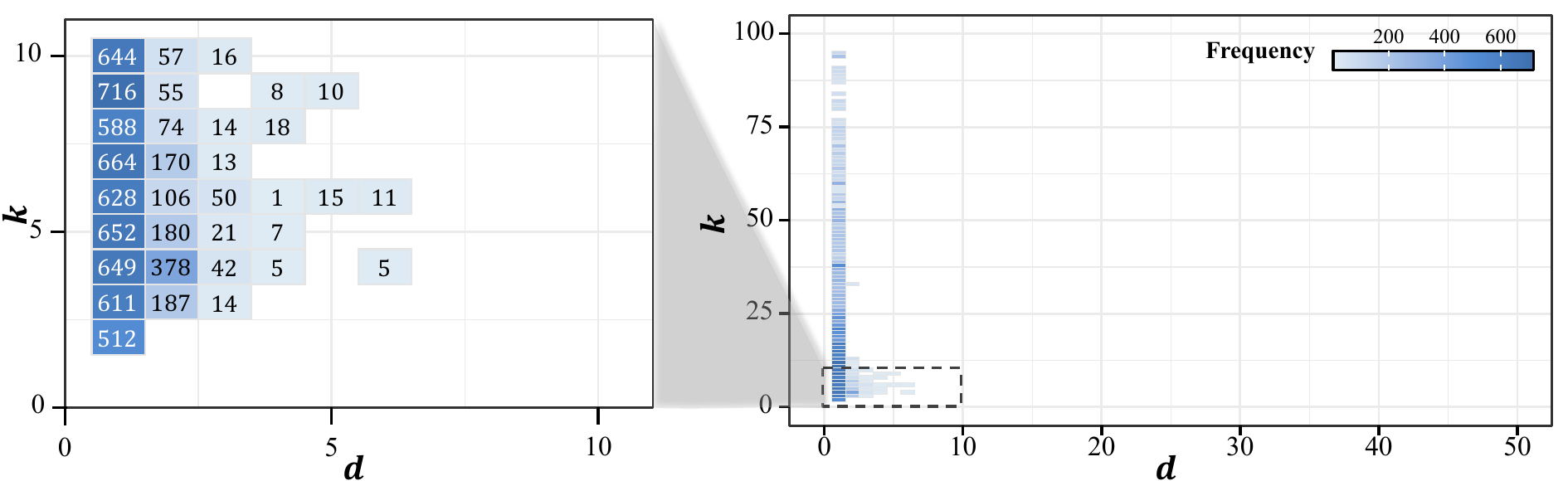}
        \vspace{-0.3cm}
        \caption{$(k,d)$-coreness}
        \label{fig:eq5_kd}
    \end{subfigure}
    \begin{subfigure}[h]{0.90\linewidth} 
        \centering
        \includegraphics[width=0.99\linewidth]{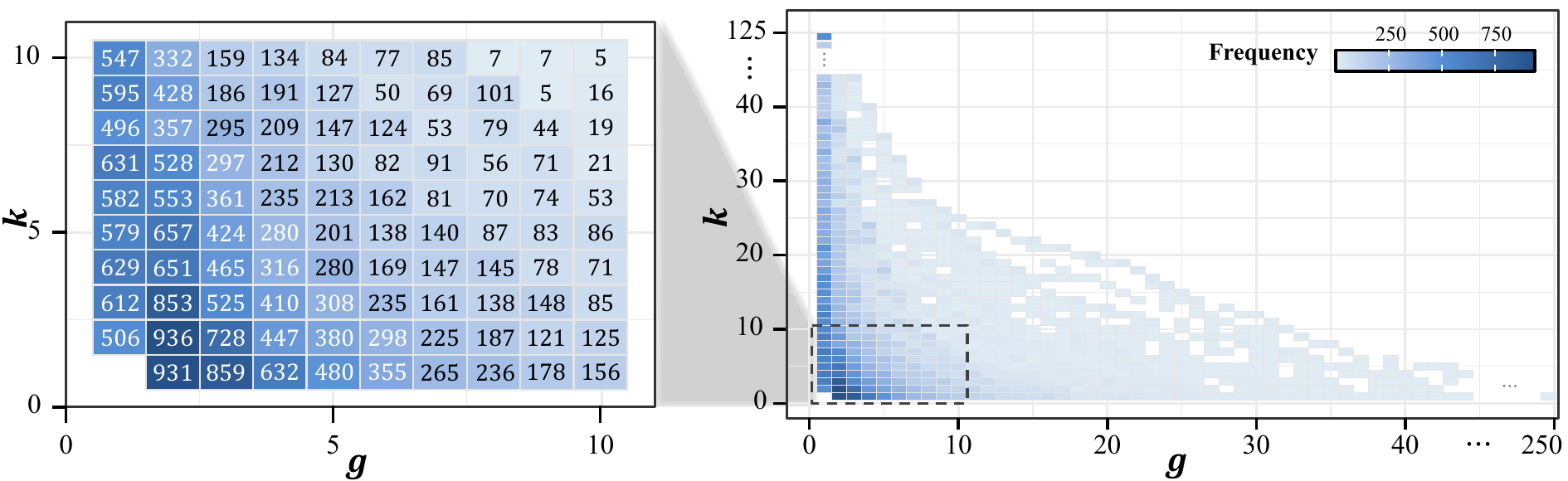}
        \vspace{-0.3cm}
        \caption{$(k,g)$-coreness}
        \label{fig:eq5_kg}
    \end{subfigure}
    \vspace{-0.2cm}
    \caption{EQ7. Distribution of $(k,g)$-Coreness and $(k,d)$-Coreness.}
    \vspace{-0.2cm}
    \label{fig:eq5}
\end{figure*}

\spara{EQ5-3. Comparison of $(k,g)$-core with Other Models (Average Support).} We further compare average support, which quantifies how frequently node pairs co-occur in hyperedges. Figure~\ref{fig:eq7_sup} shows that $(k,g)$-core model consistently yields higher average support than other models. While degree-based models such as the $(\alpha,\beta)$-core, $k$-hypercore, and $(k,d)$-core also exhibit moderately high support, their constraints focus on individual node participation rather than pairwise cohesion. Since node degree only reflects the number of hyperedges a node belongs to, it does not necessarily imply that nodes frequently co-occur. In contrast, the $(k,g)$-core explicitly enforces frequent co-occurrence, leading to the discovery of structurally cohesive subhypergraphs.

\spara{EQ6. Running Time of {\BCA} on Various Datasets.} Figure~\ref{fig:eq6_1} and Figure~\ref{fig:eq6_2} present the runtime of {\BCA} on real-world and synthetic datasets, respectively. For real-world datasets, we observe that runtime increases with the number of nodes and hyperedges, as higher values of $k$ and $g$ demand deeper decomposition. Notably, {\BCA} processes the AMiner dataset, which contains over $27$ million nodes and $17$ million hyperedges, in approximately $260,000$ seconds, demonstrating its ability to handle large-scale data. On synthetic datasets, the runtime scales nearly linearly with the number of nodes, confirming that {\BCA} is scalable even as hypergraph size grows.

\spara{EQ7. Distribution of $(k,g)$-Coreness and $(k,d)$-Coreness.} 
We compare coreness distributions produced by the $(k,g)$-core and $(k,d)$-core models, as shown in Figures~\ref{fig:eq5_kd} and \ref{fig:eq5_kg}. The $(k,g)$-core spans a wider and more fine-grained range, with values reaching $k=121$ and $g=250$, whereas the $(k,d)$-core typically peaks at $k=95$ and $d=6$, with most nodes concentrated in low $d$ values. This contrast reflects the stricter induced subgraph constraint in the $(k,d)$-core, which results in a less expressive decomposition. In contrast, the $(k,g)$-core enables more granular hierarchies, capturing a broader spectrum of internal cohesion levels.


\begin{figure*}[t]
    \centering
    \includegraphics[width=0.99\linewidth]{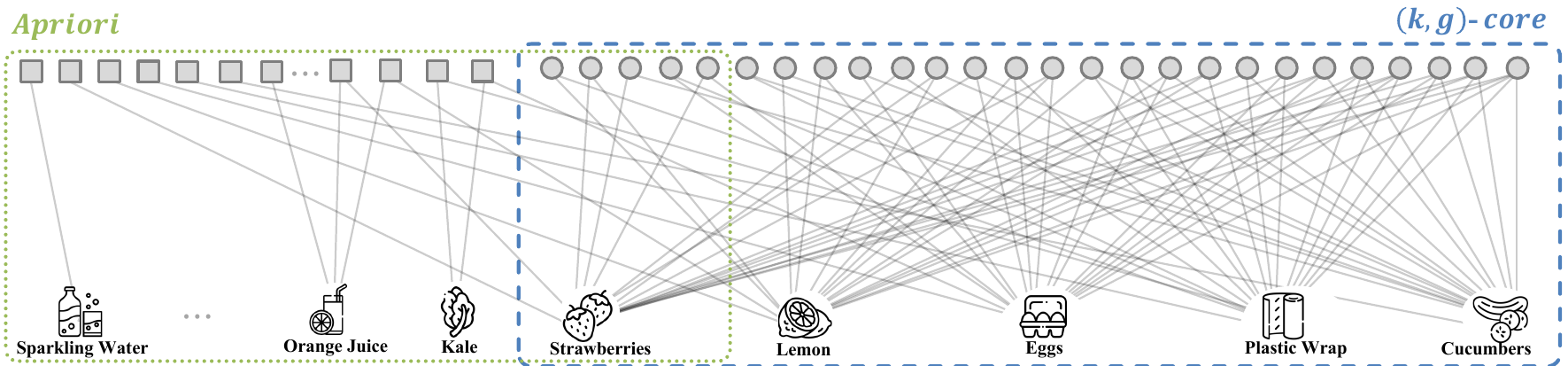}
        \vspace{-0.2cm}
        \caption{EQ8. Case study : Comparison of frequent pattern mining results between Apriori and $(k,g)$-core on the Instacart dataset.}
        \label{fig:kg_freq}
\end{figure*}

\spara{EQ8. Case Study : Frequent Pattern Mining.} We evaluate the $(k,g)$-core model against the Apriori~\cite{agrawal1994fast} algorithm using the Instacart~\cite{Benson-2018-subset} dataset, which includes over $49,000$ items and $3.3$ million transactions. With parameters $k=3$ and $g=30,000$, we set support threshold of Apriori as $30,000$. Figure~\ref{fig:kg_freq} presents both results as bipartite graphs connecting itemsets and items. The $(k,g)$-core identifies larger and denser itemsets, such as \{strawberries, lemon, cucumbers\}, while Apriori mostly returns single items or pairs, with over 85\% of its results being singletons. These findings demonstrate that $(k,g)$-core offers a more flexible and concise approach to frequent pattern mining, enabling the extraction of meaningful bundles that traditional methods may overlook.

%% file: 06_related_work.tex
\section{RELATED WORK}\label{sec:relatedwork}

\begin{table*}[!b]
\centering
\caption{Comparison of cohesive subgraph models based on constraint types}
\label{tab:model_comparison}
\begin{tabular}{c||c|c|c|c|c|c|c}
\hline \hline
\multirow{2}{*}{\textbf{Model}} & \multirow{2}{*}{\textbf{Neighbour size}} & \multirow{2}{*}{\textbf{Degree}} & \multirow{2}{*}{\textbf{Cardinality}} & \multirow{2}{*}{\textbf{Fraction}} & \multirow{2}{*}{\textbf{Support}} & \textbf{Partial} & \multirow{2}{*}{\textbf{Type}} \\ 
& & & & & & \textbf{participation}&  \\ \hline \hline
$k$-hypercore~\cite{leng2013m} & -- & $\checkmark$ & -- & -- & -- & -- & Hypergraph \\ \hline
$(k,t)$-hypercore~\cite{bu2023hypercore} & -- & $\checkmark$ & -- & $\checkmark$ & -- & $\checkmark$ & Hypergraph \\ \hline
nbr-$k$-core~\cite{nbrkcore} & $\checkmark$ & -- & -- & -- & -- & -- & Hypergraph \\ \hline
$(k,d)$-core~\cite{nbrkcore} & $\checkmark$ & $\checkmark$ & -- & -- & -- & -- & Hypergraph \\ \hline
Clique-core~\cite{batagelj2011fast} & $\checkmark$ & -- & -- & -- & -- & $\checkmark$ & Unipartite \\ \hline
$(\alpha,\beta)$-core~\cite{alphabeta} & -- & $\checkmark$ & $\checkmark$ & -- & -- & $\checkmark$ & Bipartite \\ \hline
$(k,g)$-core (our) & $\checkmark$ & -- & -- & -- & $\checkmark$ & $\checkmark$ & Hypergraph \\
\hline \hline
\end{tabular}
\end{table*}

\subsection{Cohesive Subgraph Models in Hypergraphs}

This section explores various cohesive subgraph models in hypergraphs, addressing different aspects of complex relationships. 
The $k$-hypercore~\cite{leng2013m}, as the first extension of the $k$-core~\cite{seidman1983network} for hypergraphs, ensures that each node has a minimum degree but lacks the ability to capture high-order relationships due to its limited constraint. 
To overcome this limitation, the $(k,t)$-hypercore~\cite{bu2023hypercore} introduces a proportional threshold for hyperedge inclusion. However, it does not account for the distinct roles or contributions of individual nodes within the same hyperedges, thereby failing to fully capture complex, high-order relationships among nodes. 
The nbr-$k$-core~\cite{nbrkcore} introduces a ``strongly induced subhypergraph" where a hyperedge exists if and only if all its nodes are within that subhypergraph. This model overcomes the tendency of $k$-hypercores to contain large hyperedges through entire neighbourhood inclusion, but it still struggles with large hyperedges and strict constraints. 
The $(k,d)$-core~\cite{nbrkcore} adds a degree constraint to mitigate the limitation of nbr-$k$-core, but the strict constraints may still exclude significant cohesive structures. 
The Clique-core~\cite{batagelj2011fast} transforms hypergraphs into traditional graphs, identifying $k$-cores but potentially weakening the rich interconnectivity of hypergraphs. 
Lastly, the $(\alpha, \beta)$-core~\cite{alphabeta} models cohesive structures in bipartite graphs for dual relationship analysis, but both models suffer from size inflation when converting hypergraphs~\cite{huang2015scalable}. These models offer various approaches to understanding the structure of hypergraphs but often fall short in capturing the full complexity of relationships. 

A comparative summary of the above cohesive subgraph models with our proposed $(k,g)$-core is presented in Table~\ref{tab:model_comparison}. 
Although the $(k,g)$-core does not explicitly impose structural constraints on node degree, hyperedge cardinality, or fractional participation thresholds, these constraints can be flexibly incorporated through appropriate algorithmic extensions.

\subsection{Cohesive Subgraph Models in Graphs}
In simple graphs, several cohesive subgraph models have been developed to capture network structure. The $k$-core~\cite{seidman1983network}, one of the earliest models, identifies a subgraph where each node has at least $k$ neighbours, offering a basic but efficient approach for detecting cohesive regions. This method has been extended in various directions to address its limitations. The $k$-truss~\cite{cohen2008trusses} refines cohesion by requiring that each edge be part of at least $k-2$ triangles, providing a more robust structure but resulting in smaller subgraphs. The $k$-peak~\cite{govindan2017k} seeks to identify globally cohesive subgraphs by overcoming the tendency of $k$-core to overestimate cohesive structure, ensuring that it captures only the most cohesive regions without including loosely connected parts. Models like $k$-distance clique~\cite{mokken1979cliques, wasserman1994social} generalise the concept of cliques by allowing nodes within a bounded distance to be considered cohesive. Flexi-clique~\cite{kim2024flexi} is a degree-based relaxation mode, where the minimum degree constraint increases sub-linearly with the size of the subgraph, allowing for the identification of large cohesive structures even in sparse networks.
The $k$-VCC~\cite{moody2002social} and $k$-ECC~\cite{wang2015simple} models impose vertex and edge connectivity constraints to ensure robustness against the removal of nodes or edges. While these models offer diverse perspectives on graph cohesion, they each involve trade-offs in terms of complexity, interpretability, and computational efficiency~\cite{kim2024experimental}.


%% file: 07_conclusion.tex
\section{CONCLUSION}
\label{sec:conclusion}

In this paper, we introduced the $(k,g)$-core model for cohesive subgraph discovery in hypergraphs, which better captures the relationships between nodes by incorporating both node connectivity and interaction frequency constraints. We proposed the Efficient Peeling Algorithm (\EPA) for computing the $(k,g)$-core, and the Bucket-based Coreness Algorithm (\BCA) to identify all possible $(k,g)$-cores in the network. Extensive experiments validated that our proposed algorithm is scalable and capable of handling complex, large-scale hypergraphs. 
Moreover, our model offers valuable insights in practical applications such as team formation and market basket analysis, where uncovering groups with strong internal cohesiveness and frequent co-occurrences is essential. Future work will explore extensions of the $(k,g)$-core model for dynamic and streaming environments, allowing for real-time updates as hypergraphs evolve. Additionally, there is a need to design a more space-efficient indexing structure and explore strategies to reduce time complexity, both of which are critical for improving the scalability of the model for handling large and complex datasets.

%% file: 80_appendix.tex
\section{Na\"ive Algorithm for $(k,g)$-core computation}
\label{appendix:naive}

\begin{algorithm}[h]
\SetAlgoLined
\small
\SetKwData{break}{break}
\SetKwData{return}{return}
\SetKwData{true}{true}
\SetKwData{false}{false}
\SetKwFunction{degree}{degree}
\SetKwData{del}{delete}
\SetKwFunction{getKeys}{getKeys}
\SetKwFunction{containsKey}{containsKey}
\SetKwFunction{node}{node}
\SetKwFunction{push}{enqueue}
\SetKwFunction{insert}{insert}
\SetKwFunction{pop}{dequeue}
\SetKwFunction{queueInit}{queue}
\SetKwFunction{emptyc}{empty}
\KwIn{Hypergraph $G=(V,E)$, parameters $k$ and $g$}
\KwOut{The $(k,g)$-core of $G$}
$M \leftarrow \{\}$ \; \tcp{Store all $g$-neighbours for each node}  
$H \leftarrow V;$ \tcp{Candidate nodes for $(k,g)$-core} 
$VQ \leftarrow$ \queueInit{}; \tcp{Store nodes that do not satisfy the constraints} 
\ForEach{$v \in H$}{
    $M_v \leftarrow \{\}$\;
    \ForEach{$e\in \mathcal{E}(v)$}{
        \ForEach{$u \in$ \node{$e$}}{
            \If{$u \neq v$}{
                \If{$M_v$.\containsKey{$u$} $\neq$ \true}{
                    $M_v$.\insert{$u,0$}\;
                }
                \Else{
                $M_v[u] \leftarrow M_v[u]+1$\;}
            }
        }
    }
    \ForEach{$u \in$ \getKeys{$M_v$}}{
        \If{$M_v[u] < g$}{
            \del $M_v[u]$ \;
        }
    }
    \If{$|M_v|<k$}{
        $VQ.\push{v}$\;
    }
}
\While{$VQ$.\emptyc{} $\neq$ \true}{
    $v \leftarrow VQ$.\pop{}\; 
    $H \leftarrow H \setminus \{v\}$\;
    \ForEach{$u \in$ \getKeys{$M_v$}}{
        \del $M_u[v]$\;
        \If{$|M_u| < k$}{
            $VQ.$\push{$u$}\;
        }
    }   
    \del $M_v$\; 
}
\Return{$H$}
\caption{\mbox{Na\"ive $(k,g)$-core computation~\cite{kgcore}}}
\label{alg:npa}
\end{algorithm}

The Nai\"ve algorithm, shown in Algorithm~\ref{alg:npa}, generalises the idea of iterative node removal in $k$-core computation~\cite{batagelj2003m} to the $(k,g)$-core setting. The key idea is to maintain a data structure that records, for each node $v$, the set of nodes that co-occur with $v$ in at least $g$ hyperedges (\textit{i.e.}, the $g$-neighbours of $v$). Initially, for each node $v$, we count the number of hyperedges it shares with its neighbours. After filtering out any neighbours that do not appear at least $g$ times with $v$, the remaining nodes are regarded as the $g$-neighbours. We then repeatedly remove any node $v$ that has fewer than $k$ $g$-neighbours. After each removal, the $g$-neighbour informations of the remaining nodes are updated accordingly. This peeling process continues until no more nodes can be discarded, and the remaining subhypergraph forms the maximal $(k,g)$-core.

\spara{Time complexity.}
The time complexity of the Nai\"ve algorithm has the same time complexity as {\EPA}. For every node $v$, the algorithm traverses at most $deg(v)$ hyperedges containing $v$ to count the number of co-occurrences of $v$ with its neighbours. Since a single hyperedge contains at most $|e^*|$ nodes, the time required to process $g$ neighbours for a single node is $O(\deg(v) \cdot |e^*|)$. Adding this cost up for all nodes, the overall time complexity is $O(|e^*| \cdot \mathcal{D})$, where $\mathcal{D}$ is the sum of the degrees of all nodes.

\spara{Space complexity.} 
The space complexity of the Nai\"ve algorithm is determined by the $g$-neighbour storage structure. This stores the number of shared hyperedges for all neighbour pairs of nodes. In the worst case, every node can share a hyperedge with every other node, resulting in $O(|V|^2)$. Therefore, the overall space complexity is $O(|V|^2)$.

\section{Comparison of memory usage with the Nai\"ve algorithm }
\label{appendix:naive_memory}
 \begin{figure}[h]
    \centering
    \includegraphics[width=0.99\linewidth]{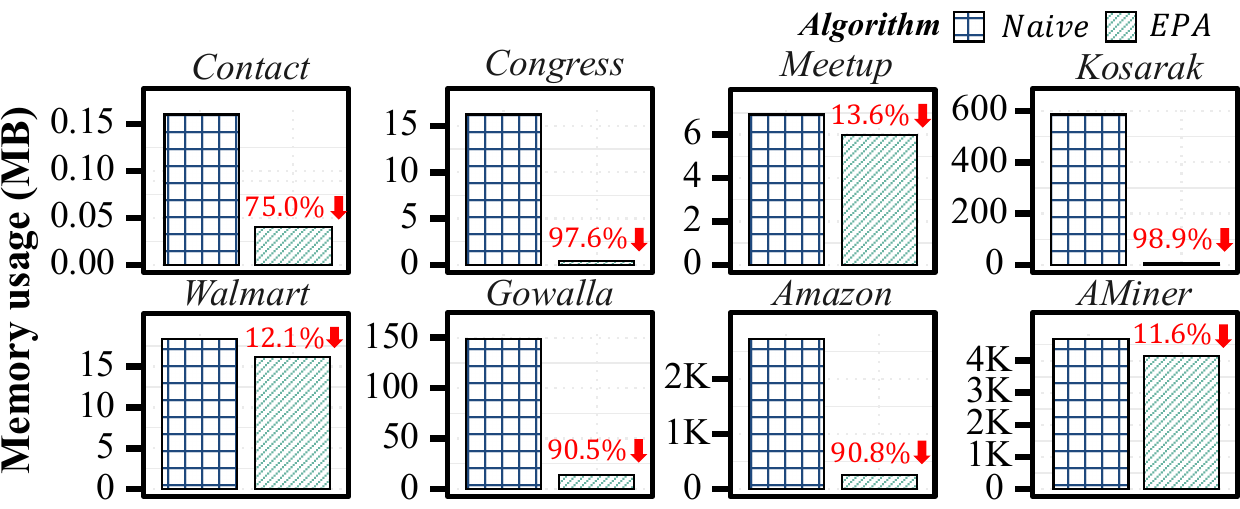}
    \vspace{-0.2cm}
    \caption{Comparions with the Nai\"ve algorithm}
    \vspace{-0.2cm}
    \label{fig:naive_memory}
 \end{figure}
Figure~\ref{fig:naive_memory} presents the peak memory consumption of the Nai\"ve algorithm and {\EPA} across various datasets in Table~\ref{tab:data}, with the memory reduction rates indicated in red-coloured text. The major difference lies in the memory storage structure for $g$-neighbours. 

\begin{itemize}[leftmargin=*]
    \item The Naïve algorithm maintains explicit storage of all $g$-neighbours for each node, leading to significant memory overhead.
    \item In contrast, {\EPA} only tracks the number of $g$-neighbours per node, substantially reducing memory requirements.
\end{itemize}

As a result, {\EPA} consistently presents lower peak memory usage across all datasets. This difference is particularly significant in datasets with large average neighbourhood sizes, such as Kosarak, Congress, Gowalla, and Amazon, where storing explicit $g$-neighbour relationships incurs considerable memory overhead. Conversely, datasets with smaller neighbourhood sizes (e.g., Meetup, Walmart, and Aminer) exhibit a lower reduction in memory usage since fewer neighbour relationships need to be stored.
In fact, the average number of $g$-neighbours in the Kosarak dataset is $709.40$, while Walmart has only $5.95$.
Overall, this efficient memory usage makes {\EPA} more suitable for large-scale hypergraphs, where managing memory becomes a critical concern for scalability and performance.


%% file: 00_main.bbl

\begin{thebibliography}{43}


\ifx \showCODEN    \undefined \def \showCODEN     #1{\unskip}     \fi
\ifx \showISBNx    \undefined \def \showISBNx     #1{\unskip}     \fi
\ifx \showISBNxiii \undefined \def \showISBNxiii  #1{\unskip}     \fi
\ifx \showISSN     \undefined \def \showISSN      #1{\unskip}     \fi
\ifx \showLCCN     \undefined \def \showLCCN      #1{\unskip}     \fi
\ifx \shownote     \undefined \def \shownote      #1{#1}          \fi
\ifx \showarticletitle \undefined \def \showarticletitle #1{#1}   \fi
\ifx \showURL      \undefined \def \showURL       {\relax}        \fi
\providecommand\bibfield[2]{#2}
\providecommand\bibinfo[2]{#2}
\providecommand\natexlab[1]{#1}
\providecommand\showeprint[2][]{arXiv:#2}

\bibitem[Agrawal et~al\mbox{.}(1994)]%
        {agrawal1994fast}
\bibfield{author}{\bibinfo{person}{Rakesh Agrawal}, \bibinfo{person}{Ramakrishnan Srikant}, {et~al\mbox{.}}} \bibinfo{year}{1994}\natexlab{}.
\newblock \showarticletitle{Fast algorithms for mining association rules}. In \bibinfo{booktitle}{\emph{Proceedings of the VLDB Endowment}}, Vol.~\bibinfo{volume}{1215}. Santiago, \bibinfo{pages}{487--499}.
\newblock


\bibitem[Amburg et~al\mbox{.}(2020)]%
        {Amburg2020categorical}
\bibfield{author}{\bibinfo{person}{Ilya Amburg}, \bibinfo{person}{Nate Veldt}, {and} \bibinfo{person}{Austin~R. Benson}.} \bibinfo{year}{2020}\natexlab{}.
\newblock \showarticletitle{Clustering in graphs and hypergraphs with categorical edge labels}. In \bibinfo{booktitle}{\emph{Proceedings of the Web Conference}}.
\newblock


\bibitem[Arafat et~al\mbox{.}(2023)]%
        {nbrkcore}
\bibfield{author}{\bibinfo{person}{Naheed~Anjum Arafat}, \bibinfo{person}{Arijit Khan}, \bibinfo{person}{Arpit~Kumar Rai}, {and} \bibinfo{person}{Bishwamittra Ghosh}.} \bibinfo{year}{2023}\natexlab{}.
\newblock \showarticletitle{Neighborhood-Based Hypergraph Core Decomposition}.
\newblock \bibinfo{journal}{\emph{Proceedings of the VLDB Endowment}} \bibinfo{volume}{16}, \bibinfo{number}{9} (\bibinfo{year}{2023}), \bibinfo{pages}{2061--2074}.
\newblock


\bibitem[Batagelj and Zaversnik(2003)]%
        {batagelj2003m}
\bibfield{author}{\bibinfo{person}{Vladimir Batagelj} {and} \bibinfo{person}{Matjaz Zaversnik}.} \bibinfo{year}{2003}\natexlab{}.
\newblock \showarticletitle{An o (m) algorithm for cores decomposition of networks}.
\newblock \bibinfo{journal}{\emph{arXiv preprint cs/0310049}} (\bibinfo{year}{2003}).
\newblock


\bibitem[Batagelj and Zaver{\v{s}}nik(2011)]%
        {batagelj2011fast}
\bibfield{author}{\bibinfo{person}{Vladimir Batagelj} {and} \bibinfo{person}{Matja{\v{z}} Zaver{\v{s}}nik}.} \bibinfo{year}{2011}\natexlab{}.
\newblock \showarticletitle{Fast algorithms for determining (generalized) core groups in social networks}.
\newblock \bibinfo{journal}{\emph{Advances in Data Analysis and Classification}} \bibinfo{volume}{5}, \bibinfo{number}{2} (\bibinfo{year}{2011}), \bibinfo{pages}{129--145}.
\newblock


\bibitem[Benson et~al\mbox{.}(2018a)]%
        {benson2018simplicial}
\bibfield{author}{\bibinfo{person}{Austin~R Benson}, \bibinfo{person}{Rediet Abebe}, \bibinfo{person}{Michael~T Schaub}, \bibinfo{person}{Ali Jadbabaie}, {and} \bibinfo{person}{Jon Kleinberg}.} \bibinfo{year}{2018}\natexlab{a}.
\newblock \showarticletitle{Simplicial closure and higher-order link prediction}.
\newblock \bibinfo{journal}{\emph{Proceedings of the National Academy of Sciences}} \bibinfo{volume}{115}, \bibinfo{number}{48} (\bibinfo{year}{2018}), \bibinfo{pages}{E11221--E11230}.
\newblock


\bibitem[Benson et~al\mbox{.}(2018b)]%
        {Benson-2018-subset}
\bibfield{author}{\bibinfo{person}{Austin~R Benson}, \bibinfo{person}{Ravi Kumar}, {and} \bibinfo{person}{Andrew Tomkins}.} \bibinfo{year}{2018}\natexlab{b}.
\newblock \showarticletitle{A discrete choice model for subset selection}. In \bibinfo{booktitle}{\emph{Proceedings of the ACM International Conference on Web Search and Data Mining}}. \bibinfo{publisher}{Association for Computing Machinery}, \bibinfo{address}{New York, NY, United States}, \bibinfo{pages}{37--45}.
\newblock


\bibitem[Berge(1984)]%
        {berge1984hypergraphs}
\bibfield{author}{\bibinfo{person}{Claude Berge}.} \bibinfo{year}{1984}\natexlab{}.
\newblock \bibinfo{booktitle}{\emph{Hypergraphs: combinatorics of finite sets}}. Vol.~\bibinfo{volume}{45}.
\newblock \bibinfo{publisher}{Elsevier}.
\newblock


\bibitem[Bu et~al\mbox{.}(2023)]%
        {bu2023hypercore}
\bibfield{author}{\bibinfo{person}{Fanchen Bu}, \bibinfo{person}{Geon Lee}, {and} \bibinfo{person}{Kijung Shin}.} \bibinfo{year}{2023}\natexlab{}.
\newblock \showarticletitle{Hypercore decomposition for non-fragile hyperedges: concepts, algorithms, observations, and applications}.
\newblock \bibinfo{journal}{\emph{Data Mining and Knowledge Discovery}} \bibinfo{volume}{37}, \bibinfo{number}{6} (\bibinfo{year}{2023}), \bibinfo{pages}{2389--2437}.
\newblock


\bibitem[Cho et~al\mbox{.}(2011)]%
        {cho2011friendship}
\bibfield{author}{\bibinfo{person}{Eunjoon Cho}, \bibinfo{person}{Seth~A Myers}, {and} \bibinfo{person}{Jure Leskovec}.} \bibinfo{year}{2011}\natexlab{}.
\newblock \showarticletitle{Friendship and mobility: user movement in location-based social networks}. In \bibinfo{booktitle}{\emph{Proceedings of the ACM SIGKDD Conference on Knowledge Discovery and Data Mining}}. \bibinfo{publisher}{Association for Computing Machinery}, \bibinfo{address}{New York, NY, USA}, \bibinfo{pages}{1082--1090}.
\newblock


\bibitem[Chu et~al\mbox{.}(2020)]%
        {chu2020finding}
\bibfield{author}{\bibinfo{person}{Deming Chu}, \bibinfo{person}{Fan Zhang}, \bibinfo{person}{Xuemin Lin}, \bibinfo{person}{Wenjie Zhang}, \bibinfo{person}{Ying Zhang}, \bibinfo{person}{Yinglong Xia}, {and} \bibinfo{person}{Chenyi Zhang}.} \bibinfo{year}{2020}\natexlab{}.
\newblock \showarticletitle{Finding the best $k$ in core decomposition: A time and space optimal solution}. In \bibinfo{booktitle}{\emph{IEEE International Conference on Data Engineering}}. \bibinfo{pages}{685--696}.
\newblock


\bibitem[{\c{C}}i{\c{c}}ekli and Kabasakal(2021)]%
        {cciccekli2021market}
\bibfield{author}{\bibinfo{person}{Ural~G{\"o}kay {\c{C}}i{\c{c}}ekli} {and} \bibinfo{person}{{\.I}nan{\c{c}} Kabasakal}.} \bibinfo{year}{2021}\natexlab{}.
\newblock \showarticletitle{Market basket analysis of basket data with demographics: a case study in e-retailing}.
\newblock \bibinfo{journal}{\emph{Alphanumeric Journal}} \bibinfo{volume}{9}, \bibinfo{number}{1} (\bibinfo{year}{2021}), \bibinfo{pages}{1--12}.
\newblock


\bibitem[Cohen(2008)]%
        {cohen2008trusses}
\bibfield{author}{\bibinfo{person}{Jonathan Cohen}.} \bibinfo{year}{2008}\natexlab{}.
\newblock \showarticletitle{Trusses: Cohesive subgraphs for social network analysis}.
\newblock \bibinfo{journal}{\emph{National security agency technical report}} \bibinfo{volume}{16}, \bibinfo{number}{3.1} (\bibinfo{year}{2008}), \bibinfo{pages}{1--29}.
\newblock


\bibitem[Ding et~al\mbox{.}(2017)]%
        {alphabeta}
\bibfield{author}{\bibinfo{person}{Danhao Ding}, \bibinfo{person}{Hui Li}, \bibinfo{person}{Zhipeng Huang}, {and} \bibinfo{person}{Nikos Mamoulis}.} \bibinfo{year}{2017}\natexlab{}.
\newblock \showarticletitle{Efficient fault-tolerant group recommendation using alpha-beta-core}. In \bibinfo{booktitle}{\emph{Proceedings of the ACM International Conference on Information and Knowledge Management}}. \bibinfo{publisher}{Association for Computing Machinery}, \bibinfo{address}{New York, NY, USA}, \bibinfo{pages}{2047--2050}.
\newblock


\bibitem[Gao et~al\mbox{.}(2022)]%
        {gao2022hgnn+}
\bibfield{author}{\bibinfo{person}{Yue Gao}, \bibinfo{person}{Yifan Feng}, \bibinfo{person}{Shuyi Ji}, {and} \bibinfo{person}{Rongrong Ji}.} \bibinfo{year}{2022}\natexlab{}.
\newblock \showarticletitle{Hgnn+: General hypergraph neural networks}.
\newblock \bibinfo{journal}{\emph{IEEE Transactions on Pattern Analysis and Machine Intelligence}} \bibinfo{volume}{45}, \bibinfo{number}{3} (\bibinfo{year}{2022}), \bibinfo{pages}{3181--3199}.
\newblock


\bibitem[Girvan and Newman(2002)]%
        {girvan2002community}
\bibfield{author}{\bibinfo{person}{Michelle Girvan} {and} \bibinfo{person}{Mark~EJ Newman}.} \bibinfo{year}{2002}\natexlab{}.
\newblock \showarticletitle{Community structure in social and biological networks}.
\newblock \bibinfo{journal}{\emph{Proceedings of the national academy of sciences}} \bibinfo{volume}{99}, \bibinfo{number}{12} (\bibinfo{year}{2002}), \bibinfo{pages}{7821--7826}.
\newblock


\bibitem[Govindan et~al\mbox{.}(2017)]%
        {govindan2017k}
\bibfield{author}{\bibinfo{person}{Priya Govindan}, \bibinfo{person}{Chenghong Wang}, \bibinfo{person}{Chumeng Xu}, \bibinfo{person}{Hongyu Duan}, {and} \bibinfo{person}{Sucheta Soundarajan}.} \bibinfo{year}{2017}\natexlab{}.
\newblock \showarticletitle{The k-peak decomposition: Mapping the global structure of graphs}. In \bibinfo{booktitle}{\emph{Proceedings of the 26th International Conference on World Wide Web}}. \bibinfo{publisher}{International World Wide Web Conferences Steering Committee}, \bibinfo{address}{Republic and Canton of Geneva, Switzerland}, \bibinfo{pages}{1441--1450}.
\newblock


\bibitem[G{\"u}nnemann et~al\mbox{.}(2013)]%
        {gunnemann2013spectral}
\bibfield{author}{\bibinfo{person}{Stephan G{\"u}nnemann}, \bibinfo{person}{Ines F{\"a}rber}, \bibinfo{person}{Sebastian Raubach}, {and} \bibinfo{person}{Thomas Seidl}.} \bibinfo{year}{2013}\natexlab{}.
\newblock \showarticletitle{Spectral subspace clustering for graphs with feature vectors}. In \bibinfo{booktitle}{\emph{IEEE International Conference on Data Mining}}. IEEE, \bibinfo{publisher}{IEEE}, \bibinfo{pages}{231--240}.
\newblock


\bibitem[Harary and Norman(1953)]%
        {harary1953graph}
\bibfield{author}{\bibinfo{person}{Frank Harary} {and} \bibinfo{person}{Robert~Z Norman}.} \bibinfo{year}{1953}\natexlab{}.
\newblock \showarticletitle{Graph theory as a mathematical model in social science}.
\newblock  (\bibinfo{year}{1953}).
\newblock


\bibitem[Huang et~al\mbox{.}(2015)]%
        {huang2015scalable}
\bibfield{author}{\bibinfo{person}{Jin Huang}, \bibinfo{person}{Rui Zhang}, {and} \bibinfo{person}{Jeffrey~Xu Yu}.} \bibinfo{year}{2015}\natexlab{}.
\newblock \showarticletitle{Scalable hypergraph learning and processing}. In \bibinfo{booktitle}{\emph{IEEE International Conference on Data Mining}}. \bibinfo{publisher}{IEEE Computer Society}, \bibinfo{address}{Los Alamitos, CA, USA}, \bibinfo{pages}{775--780}.
\newblock


\bibitem[Jack et~al\mbox{.}(2013)]%
        {jack2013transaction}
\bibfield{author}{\bibinfo{person}{William Jack}, \bibinfo{person}{Adam Ray}, {and} \bibinfo{person}{Tavneet Suri}.} \bibinfo{year}{2013}\natexlab{}.
\newblock \showarticletitle{Transaction networks: Evidence from mobile money in Kenya}.
\newblock \bibinfo{journal}{\emph{American Economic Review}} \bibinfo{volume}{103}, \bibinfo{number}{3} (\bibinfo{year}{2013}), \bibinfo{pages}{356--361}.
\newblock


\bibitem[Kami{\'n}ski et~al\mbox{.}(2023)]%
        {kaminski2023hypergraph}
\bibfield{author}{\bibinfo{person}{Bogumi{\l} Kami{\'n}ski}, \bibinfo{person}{Pawe{\l} Pra{\l}at}, {and} \bibinfo{person}{Fran{\c{c}}ois Th{\'e}berge}.} \bibinfo{year}{2023}\natexlab{}.
\newblock \showarticletitle{Hypergraph Artificial Benchmark for Community Detection (h--ABCD)}.
\newblock \bibinfo{journal}{\emph{Journal of Complex Networks}} \bibinfo{volume}{11}, \bibinfo{number}{4} (\bibinfo{year}{2023}).
\newblock


\bibitem[Kim et~al\mbox{.}(2023)]%
        {kgcore}
\bibfield{author}{\bibinfo{person}{Dahee Kim}, \bibinfo{person}{Junghoon Kim}, \bibinfo{person}{Sungsu Lim}, {and} \bibinfo{person}{Hyun~Ji Jeong}.} \bibinfo{year}{2023}\natexlab{}.
\newblock \showarticletitle{Exploring Cohesive Subgraphs in Hypergraphs: The $(k,g)$-core Approach}. In \bibinfo{booktitle}{\emph{Proceedings of the ACM International Conference on Information and Knowledge Management}}. \bibinfo{publisher}{Association for Computing Machinery}, \bibinfo{address}{New York, NY, USA}, \bibinfo{pages}{4013--4017}.
\newblock


\bibitem[Kim et~al\mbox{.}(2024a)]%
        {kim2024experimental}
\bibfield{author}{\bibinfo{person}{Dahee Kim}, \bibinfo{person}{Song Kim}, \bibinfo{person}{Jeongseon Kim}, \bibinfo{person}{Junghoon Kim}, \bibinfo{person}{Kaiyu Feng}, \bibinfo{person}{Sungsu Lim}, {and} \bibinfo{person}{Jungeun Kim}.} \bibinfo{year}{2024}\natexlab{a}.
\newblock \showarticletitle{Experimental Analysis and Evaluation of Cohesive Subgraph Discovery}.
\newblock \bibinfo{journal}{\emph{Information Sciences}}  \bibinfo{volume}{672} (\bibinfo{year}{2024}), \bibinfo{pages}{120664}.
\newblock


\bibitem[Kim et~al\mbox{.}(2024b)]%
        {kim2024flexi}
\bibfield{author}{\bibinfo{person}{Song Kim}, \bibinfo{person}{Junghoon Kim}, \bibinfo{person}{Susik Yoon}, {and} \bibinfo{person}{Jungeun Kim}.} \bibinfo{year}{2024}\natexlab{b}.
\newblock \showarticletitle{Flexi-clique: Exploring Flexible and Sub-linear Clique Structures}. In \bibinfo{booktitle}{\emph{Proceedings of the ACM International Conference on Information and Knowledge Management}}. \bibinfo{pages}{3832--3836}.
\newblock


\bibitem[Leng et~al\mbox{.}(2013)]%
        {leng2013m}
\bibfield{author}{\bibinfo{person}{Ming Leng}, \bibinfo{person}{Lingyu Sun}, \bibinfo{person}{Ji-nian Bian}, {and} \bibinfo{person}{Yuchun Ma}.} \bibinfo{year}{2013}\natexlab{}.
\newblock \showarticletitle{An o(m) algorithm for cores decomposition of undirected hypergraph}.
\newblock \bibinfo{journal}{\emph{Journal of Chinese Computer Systems}} \bibinfo{volume}{34}, \bibinfo{number}{11} (\bibinfo{year}{2013}), \bibinfo{pages}{2568--2573}.
\newblock


\bibitem[Li and Shan(2010)]%
        {li2010team}
\bibfield{author}{\bibinfo{person}{Cheng-Te Li} {and} \bibinfo{person}{Man-Kwan Shan}.} \bibinfo{year}{2010}\natexlab{}.
\newblock \showarticletitle{Team formation for generalized tasks in expertise social networks}. In \bibinfo{booktitle}{\emph{IEEE second international conference on social computing}}. IEEE, \bibinfo{pages}{9--16}.
\newblock


\bibitem[Li et~al\mbox{.}(2023)]%
        {li2023hypergraph}
\bibfield{author}{\bibinfo{person}{Mengran Li}, \bibinfo{person}{Yong Zhang}, \bibinfo{person}{Xiaoyong Li}, \bibinfo{person}{Yuchen Zhang}, {and} \bibinfo{person}{Baocai Yin}.} \bibinfo{year}{2023}\natexlab{}.
\newblock \showarticletitle{Hypergraph transformer neural networks}.
\newblock \bibinfo{journal}{\emph{ACM Transactions on Knowledge Discovery from Data}} \bibinfo{volume}{17}, \bibinfo{number}{5} (\bibinfo{year}{2023}), \bibinfo{pages}{1--22}.
\newblock


\bibitem[Linghu et~al\mbox{.}(2020)]%
        {linghu2020global}
\bibfield{author}{\bibinfo{person}{Qingyuan Linghu}, \bibinfo{person}{Fan Zhang}, \bibinfo{person}{Xuemin Lin}, \bibinfo{person}{Wenjie Zhang}, {and} \bibinfo{person}{Ying Zhang}.} \bibinfo{year}{2020}\natexlab{}.
\newblock \showarticletitle{Global reinforcement of social networks: The anchored coreness problem}. In \bibinfo{booktitle}{\emph{Proceedings of the ACM SIGMOD International Conference on Management of Data}}. \bibinfo{pages}{2211--2226}.
\newblock


\bibitem[Malliaros et~al\mbox{.}(2020)]%
        {malliaros2020core}
\bibfield{author}{\bibinfo{person}{Fragkiskos~D Malliaros}, \bibinfo{person}{Christos Giatsidis}, \bibinfo{person}{Apostolos~N Papadopoulos}, {and} \bibinfo{person}{Michalis Vazirgiannis}.} \bibinfo{year}{2020}\natexlab{}.
\newblock \showarticletitle{The core decomposition of networks: Theory, algorithms and applications}.
\newblock \bibinfo{journal}{\emph{Proceedings of the VLDB Endowment}} \bibinfo{volume}{29}, \bibinfo{number}{1} (\bibinfo{year}{2020}), \bibinfo{pages}{61--92}.
\newblock


\bibitem[Mason and Verwoerd(2007)]%
        {mason2007graph}
\bibfield{author}{\bibinfo{person}{Oliver Mason} {and} \bibinfo{person}{Mark Verwoerd}.} \bibinfo{year}{2007}\natexlab{}.
\newblock \showarticletitle{Graph theory and networks in biology}.
\newblock \bibinfo{journal}{\emph{IET systems biology}} \bibinfo{volume}{1}, \bibinfo{number}{2} (\bibinfo{year}{2007}), \bibinfo{pages}{89--119}.
\newblock


\bibitem[Mokken et~al\mbox{.}(1979)]%
        {mokken1979cliques}
\bibfield{author}{\bibinfo{person}{Robert~J Mokken} {et~al\mbox{.}}} \bibinfo{year}{1979}\natexlab{}.
\newblock \showarticletitle{Cliques, clubs and clans}.
\newblock \bibinfo{journal}{\emph{Quality \& Quantity}} \bibinfo{volume}{13}, \bibinfo{number}{2} (\bibinfo{year}{1979}), \bibinfo{pages}{161--173}.
\newblock


\bibitem[Moody and White(2002)]%
        {moody2002social}
\bibfield{author}{\bibinfo{person}{James Moody} {and} \bibinfo{person}{Douglas~R White}.} \bibinfo{year}{2002}\natexlab{}.
\newblock \showarticletitle{Social cohesion and embeddedness: A hierarchical conception of social groups}.
\newblock \bibinfo{journal}{\emph{Sociological Methodology}} \bibinfo{volume}{68}, \bibinfo{number}{1} (\bibinfo{year}{2002}), \bibinfo{pages}{365--368}.
\newblock


\bibitem[Ni et~al\mbox{.}(2019)]%
        {ni2019justifying}
\bibfield{author}{\bibinfo{person}{Jianmo Ni}, \bibinfo{person}{Jiacheng Li}, {and} \bibinfo{person}{Julian McAuley}.} \bibinfo{year}{2019}\natexlab{}.
\newblock \showarticletitle{Justifying Recommendations using Distantly-Labeled Reviews and Fine-Grained Aspects}. In \bibinfo{booktitle}{\emph{Proceedings of the Conference on Empirical Methods in Natural Language Processing and the International Joint Conference on Natural Language Processing}}. \bibinfo{publisher}{Association for Computational Linguistics}, \bibinfo{address}{Hong Kong, China}, \bibinfo{pages}{188--197}.
\newblock


\bibitem[Salem et~al\mbox{.}(2012)]%
        {salem2012discovering}
\bibfield{author}{\bibinfo{person}{Saeed Salem}, \bibinfo{person}{Rami Alroobi}, \bibinfo{person}{Syed Ahmed}, {and} \bibinfo{person}{Mohammad Hossain}.} \bibinfo{year}{2012}\natexlab{}.
\newblock \showarticletitle{Discovering maximal cohesive subgraphs and patterns from attributed biological networks}. In \bibinfo{booktitle}{\emph{IEEE International Conference on Bioinformatics and Biomedicine Workshops}}. IEEE, \bibinfo{publisher}{IEEE}, \bibinfo{pages}{203--210}.
\newblock


\bibitem[Saputra et~al\mbox{.}(2023)]%
        {saputra2023market}
\bibfield{author}{\bibinfo{person}{Jeffri Prayitno~Bangkit Saputra}, \bibinfo{person}{Silvia~Anggun Rahayu}, {and} \bibinfo{person}{Taqwa Hariguna}.} \bibinfo{year}{2023}\natexlab{}.
\newblock \showarticletitle{Market basket analysis using FP-growth algorithm to design marketing strategy by determining consumer purchasing patterns}.
\newblock \bibinfo{journal}{\emph{Journal of Applied Data Sciences}} \bibinfo{volume}{4}, \bibinfo{number}{1} (\bibinfo{year}{2023}), \bibinfo{pages}{38--49}.
\newblock


\bibitem[Seidman(1983)]%
        {seidman1983network}
\bibfield{author}{\bibinfo{person}{Stephen~B Seidman}.} \bibinfo{year}{1983}\natexlab{}.
\newblock \showarticletitle{Network structure and minimum degree}.
\newblock \bibinfo{journal}{\emph{Social networks}} \bibinfo{volume}{5}, \bibinfo{number}{3} (\bibinfo{year}{1983}), \bibinfo{pages}{269--287}.
\newblock


\bibitem[{\"U}nvan(2021)]%
        {unvan2021market}
\bibfield{author}{\bibinfo{person}{Y{\"u}ksel~Akay {\"U}nvan}.} \bibinfo{year}{2021}\natexlab{}.
\newblock \showarticletitle{Market basket analysis with association rules}.
\newblock \bibinfo{journal}{\emph{Communications in Statistics-Theory and Methods}} \bibinfo{volume}{50}, \bibinfo{number}{7} (\bibinfo{year}{2021}), \bibinfo{pages}{1615--1628}.
\newblock


\bibitem[Wang et~al\mbox{.}(2015)]%
        {wang2015simple}
\bibfield{author}{\bibinfo{person}{Tianhao Wang}, \bibinfo{person}{Yong Zhang}, \bibinfo{person}{Francis~YL Chin}, \bibinfo{person}{Hing-Fung Ting}, \bibinfo{person}{Yung~H Tsin}, {and} \bibinfo{person}{Sheung-Hung Poon}.} \bibinfo{year}{2015}\natexlab{}.
\newblock \showarticletitle{A simple algorithm for finding all k-edge-connected components}.
\newblock \bibinfo{journal}{\emph{PLOS One}} \bibinfo{volume}{10}, \bibinfo{number}{9} (\bibinfo{year}{2015}), \bibinfo{pages}{e0136264}.
\newblock


\bibitem[Wasserman et~al\mbox{.}(1994)]%
        {wasserman1994social}
\bibfield{author}{\bibinfo{person}{Stanley Wasserman}, \bibinfo{person}{Katherine Faust}, {et~al\mbox{.}}} \bibinfo{year}{1994}\natexlab{}.
\newblock \bibinfo{booktitle}{\emph{Social network analysis: Methods and applications}}.
\newblock \bibinfo{publisher}{Cambridge university press}.
\newblock


\bibitem[Wellman(1983)]%
        {wellman1983network}
\bibfield{author}{\bibinfo{person}{Barry Wellman}.} \bibinfo{year}{1983}\natexlab{}.
\newblock \showarticletitle{Network analysis: Some basic principles}.
\newblock \bibinfo{journal}{\emph{Sociological theory}} (\bibinfo{year}{1983}), \bibinfo{pages}{155--200}.
\newblock


\bibitem[Xia et~al\mbox{.}(2021)]%
        {xia2021self}
\bibfield{author}{\bibinfo{person}{Xin Xia}, \bibinfo{person}{Hongzhi Yin}, \bibinfo{person}{Junliang Yu}, \bibinfo{person}{Qinyong Wang}, \bibinfo{person}{Lizhen Cui}, {and} \bibinfo{person}{Xiangliang Zhang}.} \bibinfo{year}{2021}\natexlab{}.
\newblock \showarticletitle{Self-supervised hypergraph convolutional networks for session-based recommendation}. In \bibinfo{booktitle}{\emph{Proceedings of the AAAI Conference on Artificial Intelligence}}, Vol.~\bibinfo{volume}{35}. \bibinfo{pages}{4503--4511}.
\newblock


\bibitem[Zhang et~al\mbox{.}(2023)]%
        {zhang2023quantifying}
\bibfield{author}{\bibinfo{person}{Fan Zhang}, \bibinfo{person}{Qingyuan Linghu}, \bibinfo{person}{Jiadong Xie}, \bibinfo{person}{Kai Wang}, \bibinfo{person}{Xuemin Lin}, {and} \bibinfo{person}{Wenjie Zhang}.} \bibinfo{year}{2023}\natexlab{}.
\newblock \showarticletitle{Quantifying Node Importance over Network Structural Stability}. In \bibinfo{booktitle}{\emph{Proceedings of the ACM SIGKDD Conference on Knowledge Discovery and Data Mining}}. \bibinfo{pages}{3217--3228}.
\newblock


\end{thebibliography}
